\documentclass{elsart}

\usepackage[square,comma]{natbib}
\usepackage{graphicx}
\usepackage{pxfonts}
\usepackage{lineno}

\usepackage{amssymb}

\journal{}

\begin{document}

\thispagestyle{empty}
\begin{Large}
\textbf{DEUTSCHES ELEKTRONEN-SYNCHROTRON}

\textbf{\large{Ein Forschungszentrum der
Helmholtz-Gemeinschaft}\\}
\end{Large}

DESY 11-055

April 2011

\begin{eqnarray}
\nonumber &&\cr \nonumber && \cr \nonumber &&\cr
\end{eqnarray}
\begin{eqnarray}
\nonumber
\end{eqnarray}
\begin{center}
\begin{Large}
\textbf{Gas-filled cell as a narrow bandwidth bandpass filter in the VUV wavelength
range}
\end{Large}
\begin{eqnarray}
\nonumber &&\cr \nonumber && \cr
\end{eqnarray}

\begin{large}
Gianluca Geloni,
\end{large}
\textsl{\\European XFEL GmbH, Hamburg}
\begin{large}

Vitali Kocharyan and Evgeni Saldin
\end{large}
\textsl{\\Deutsches Elektronen-Synchrotron DESY, Hamburg}
\begin{eqnarray}
\nonumber
\end{eqnarray}
\begin{eqnarray}
\nonumber
\end{eqnarray}
ISSN 0418-9833
\begin{eqnarray}
\nonumber
\end{eqnarray}
\begin{large}
\textbf{NOTKESTRASSE 85 - 22607 HAMBURG}
\end{large}
\end{center}
\clearpage
\newpage

\begin{frontmatter}



\title{Gas-filled cell as a narrow bandwidth bandpass filter in the VUV wavelength
range}


\author[XFEL]{Gianluca Geloni\thanksref{corr},}
\thanks[corr]{Corresponding Author. E-mail address: gianluca.geloni@xfel.eu}
\author[DESY]{Vitali Kocharyan}
\author[DESY]{and Evgeni Saldin}

\address[XFEL]{European XFEL GmbH, Hamburg, Germany}
\address[DESY]{Deutsches Elektronen-Synchrotron (DESY), Hamburg,
Germany}

\begin{abstract}
We propose a method for spectrally filtering radiation in the VUV wavelength range by means of a monochromator constituted by a cell filled with a resonantly absorbing rare gas. Around particular wavelengths, the gas exhibits narrow-bandwidth absorbing resonances following the Fano profile. In particular, within the photon energy range  $60$ eV - $65$ eV, the correlation index of the Fano profiles for the photo-ionization spectra in Helium is equal to unity, meaning that the minimum of the cross-section is exactly zero. For sufficiently large column density in the gas cell, the spectrum of the incoming radiation will be attenuated by the background cross-section of many orders of magnitude, except for those wavelengths close to the point where the cross-section is zero. Remarkable advantages of a gas monochromator based on this principle are simplicity, efficiency and narrow-bandwidth. A gas monochromator installed in the experimental hall of a VUV SASE FEL facility would enable the delivery of a single-mode VUV laser beam. The design is identical to that of already existing gas attenuator systems for VUV or X-ray FELs. We present feasibility study and exemplifications for the FLASH facility in the VUV regime.
\end{abstract}

%
%
%
\end{frontmatter}



\section{\label{sec:intro} Introduction}

A large portion of experiments using VUV radiation are presently carried out at synchrotron radiation facilities and VUV SASE FELs. A high-resolution VUV monochromator is an important element of such facilities. Photo-absorption spectra of Helium are often used for calibration of these monochromators. In fact, as is well-known, the transmittance spectrum in the Helium gas exhibits narrow-bandwidth absorbing resonances down to a few meV, which follow the Fano profile \cite{FANO2,MORG}.

Here we propose a novel method for spectral filtering based on a remarkable feature of the Fano interference phenomenon for Helium. In particular, within the photon energy range between $60$ eV and $65$ eV, the correlation index of the Fano profile for the photo-ionization spectra in Helium is equal to unity, meaning that the minimum of the cross-section is strictly zero \cite{MORG}.  We suggest the exploitation of this feature for filtering purposes, using a windowless gas-filled cell equipped with differential pumping. In fact, for a sufficiently large column density, the spectrum of the incoming radiation will be attenuated by the background cross section of many orders of magnitude, except for those wavelength near the point where the cross-section is zero. In this paper we concentrate on physics issues. The design features of a gas monochromator system based on this principle should not differ too much from the gas attenuator systems at XFELs  \cite{TIED,LCLS2}.

A remarkable advantage of gas monochromators based on the above principle is simplicity. In fact, such a device would use no optical components, and therefore there are no problem with alignment, heat loading and beam wavefront perturbations. Moreover, the peak efficiency of a gas monochromator would be close to $100 \%$. Also, a very narrow bandwidth would be granted. Such a device, installed at any VUV SASE FEL facility would enable the delivery of single-mode VUV laser beam. In this case, the shapes of the single-shot spectra after the monochromator can be identified with the gas-cell transmittance calculated from the first principles, and do not depend on the detailed intensity distribution of modes of the incoming spectrum.

We present a feasibility study and exemplifications for the FLASH SASE FEL in the VUV regime. The applicability of the gas monochromator setup is obviously not restricted to VUV SASE FEL facilities. Synchrotron radiation facilities may benefit from this scheme as well.

\section{\label{sec:due}  Principles of the spectral filtering technique based on the use gas cell}

\begin{figure}[tb]
\includegraphics[width=1.0\textwidth]{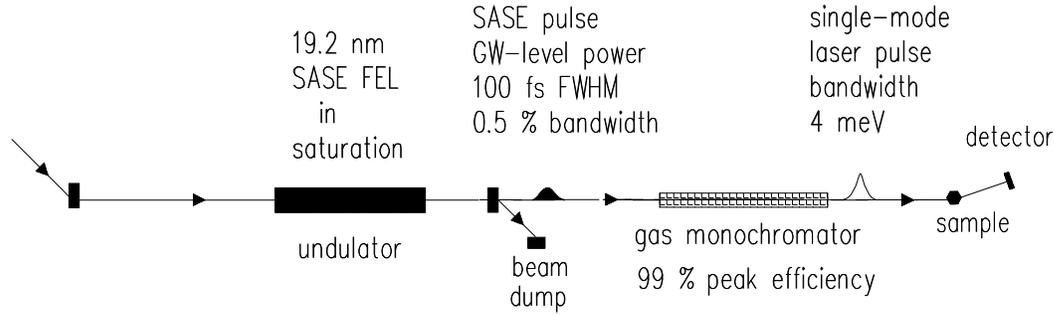}
\caption{Scheme of the gas monochromator for VUV SASE FEL sources} \label{gcell0}
\end{figure}
The output radiation pulses from SASE FEL exhibit poor longitudinal coherence. While a single-shot spectrum consists of many spikes, or modes, the average spectrum is constituted by a smooth distribution with a FWHM typically corresponding to about  $1 \%$ in the VUV wavelength range. The bandwidth of the VUV beam can be thus further reduced by using a gas monochromator. A schematic representation of the setup is shown in Fig. \ref{gcell0}. Each spike in the spectrum has an average width of about $0.03 \% - 0.04 \%$ for wavelengths around $20$ nm, and is thus larger than the bandwidth of the gas monochromator, which allows for a few meV bandwidth. This fact enables the delivery of single-mode VUV laser beams. The shape of the single shot spectra after the gas monochromator can be described in terms of the transmittance of the gas cell, and do not depend on the details of the intensity distribution of the spikes in the incoming spectrum. It is therefore possible to properly define amplitude and phase for a single-shot spectrum after the monochromator and, hence the single-shot radiation pulse shape in the time domain.

The working principle of the gas monochromator is based on the characteristic of the photo-absorption cross-sections for Helium at energies below the second ionization threshold.  In general, the interaction with a soft X-ray electromagnetic pulse has an ionizing effect for the atom.  A first ionization threshold, where the remaining electron in the ion is found in the state with principal quantum number $n=1$, is found at $24.6$ eV, while a second ionization threshold with $n=2$ is found at $65.4$ eV. Therefore, after interaction with a photon with energy between these two thresholds,  one finds the Helium ion in the ground state $1s$, and a free electron with kinetic energy equal to the photon energy diminished of  $24.6$ eV. This single open channel can be reached either directly or via decay of a doubly excited state, the well-known autoionizing state \cite{FANO2,MORG}. In this case only one channel is open in the continuum, meaning that the final state is the Helium ion in the ground state. The two paths towards this channels interfere and one is left with Fano lines obeying:

\begin{eqnarray}
\sigma = \sigma_b \frac{(q+\mathcal{E})^2}{1+\mathcal{E}^2}~,
\label{sigsig2}
\end{eqnarray}
where $\sigma_b=\sigma_b(\lambda)$ is the background cross-section, $q$ is known as the asymmetry index, while the reduced energy $\mathcal{E}$ is defined as

\begin{eqnarray}
\mathcal{E} = \frac{2(E_{R}-hc/\lambda)}{\Gamma}~,
\label{En}
\end{eqnarray}
$E_{R}$ being the energy of the resonance, with a width $\Gamma$. When the energy increases, or one considers other gases, new channels open up.
For example, above the second ionization threshold for Helium, when the photon energy increases between $65.4$ eV to $72.9$ eV, a second channel is present, characterized by  the Helium ion with the bound electron with $n=2$ ($2s$ state), and free electron energy equal to the photon energy diminished by $65.4$ eV. In this photon energy range we have a set of autoionizing resonances,which interfere with two open channels ($1s$ or $2s$). More in general, as the photon energy increases below the double ionization threshold at $79$ eV, other channels corresponding to the Helium ion with higher numbers of $n$ ($n$th ionization threshold) can be reached. Then, as explained for example in \cite{ARGO}, transitions to the continuum may or may not interact with the discrete autoionization state, and the expression for the cross-section has to be modified as:

\begin{eqnarray}
\sigma = \sigma_b \frac{(q+\mathcal{E})^2}{1+\mathcal{E}^2} + \sigma_a~,
\label{sigsig2}
\end{eqnarray}
where $\sigma_b$ and $\sigma_a$ are the background cross-sections of the interacting and non-interacting transitions to the continuum. Defining the total cross-section $\sigma_0= \sigma_a+\sigma_b$, which one would measure without autoionization, Eq. (\ref{sigsig2}) can be cast in the form

\begin{eqnarray}
\sigma = \sigma_0 \left(\rho^2 \frac{(q+\mathcal{E})^2}{1+\mathcal{E}^2} -\rho^2+1 \right)~,
\label{sigsig3}
\end{eqnarray}
where the correlation coefficient $\rho^2$ is introduced as

\begin{eqnarray}
\rho^2 = \frac{\sigma_b}{\sigma_b+\sigma_a}~.
\label{rho2}
\end{eqnarray}
Eq. (\ref{sigsig3}) reduces to Eq. (\ref{sigsig2}) for $\rho^2 =1$, corresponding to the Helium resonances under study. In this case, there is always one energy value around the resonance, at $\mathcal{E} = -q$, where $\sigma = 0$. Around that energy value, radiation can pass through the gas cell almost without attenuation. In fact, the photoabsorption cross section $\sigma$ is linked to the light attenuation through a gas medium of column density $n_0 l$, $l$ being the length of the cell and $n_0$ the gas density, via the Beer-Lambert law:

\begin{eqnarray}
I(\omega) = I_0 \exp[-n_0 l \sigma(\omega)]~,
\label{beerl}
\end{eqnarray}
where $I_0$ is the incident intensity, and $I(\omega)$ is the attenuated intensity of the transmitted light at frequency $\omega$. By inspecting Eq. (\ref{beerl}) it can be seen that when the column density is increased radiation is attenuated at all frequencies except at $\mathcal{E} = -q$. As a result, in the case of the Helium gas, and for energies below the second ionization threshold an increase in the gas density transforms the gas cell into a bandpass filter.

\subsection{\label{sec:tre}  VUV photo absorption spectra of He for photon energies between $60$ eV and $65$ eV}

Let us then consider the series of autoionizing resonances converging to the $n = 2$ level of the Helium ion, corresponding $2n+1 = 3$ Rydberg series, all converging to the second ionization threshold, corresponding to $65.4$ eV. These three series all begin with the two Helium electrons characterized by $n_1 = 2$ and $n_2 = 2$, which corresponds to an energy of $60$ eV. There are three possible choices for the $(l_1,l_2)$ quantum numbers: $(2s,np)~^1P_0$, $(2p,ns)~^1P_0$, and $(2p,nd)~^1P_0$. However, the levels of the $(2s,np)~^1P_0$ and $(2p,ns)~^1P_0$ wave functions are nearly degenerate, and the doubly-excited states are better described in terms of constructive and destructive superpositions of these wave functions, finally yielding the three Rydberg series: $(sp,2n+)~^1P_0$, $(sp,2n-)~^1P_0$, and $(2p,nd)~^1P_0$. Since these three series converge (but do not reach) the second ionization threshold, all the correspondent doubly excited states $\mathrm{He}^*$ decay to an Helium ion in the ground state. Out of these three series, only the first $(sp,2n+)$ will be of interest to us, as the others are too weak.

The energy-dependent background cross-section expressed in Megabarn ($1\mathrm{Mb} = 10^{-18} \mathrm{cm}^{2}$) is given by \cite{MORG}

\begin{eqnarray}
\sigma_b(\lambda) = -0.05504-1.3624\cdot 10^{-4} \lambda + 3.3822\cdot 10^{-5} \lambda^2~,
\label{sigb}
\end{eqnarray}
with $\lambda$ the radiation wavelength in Angstrom units. The asymmetry index $q$, the energy of the resonances $E_{R}$, and the resonance widths $\Gamma$, have been the subject of several calculations. The cross-section for the series $(sp,2n+)$ can be calculated using Eq. (\ref{sigsig2}) with the parameters given in Table \ref{tableHe}, reproduced from \cite{MORG}.

\begin{table}
  \caption{Parameters determining the Fano autoionization profiles for the  $(sp,2n+)$ autoionizing series of Helium, following \cite{MORG}.}
{\begin{tabular}[l]{@{}|c||c|c|c|c|c|c|c|c|c|}

\hline
    n              & 2      &3      &4      &5      &6      &7      &8      &9      &10          \\
    $\Gamma$  (eV) & 0.0378 & 0.0083& 0.0038& 0.0014& 0.0008& 0.0005& 0.0003& 0.0002& 0.0001     \\
    $E_{res}$ (eV) &60.14   & 63.655& 64.466& 64.816& 64.999& 65.108& 65.181& 65.229& 65.263     \\
    q              & -2.57  & -2.5  & -2.5  & -2.5  & -2.5  & -2.5  & -2.5  & - 2.5 & -2.5       \\
\hline
  \end{tabular}}
  \label{tableHe}
\end{table}

\begin{figure}
\includegraphics[width=1.0\textwidth]{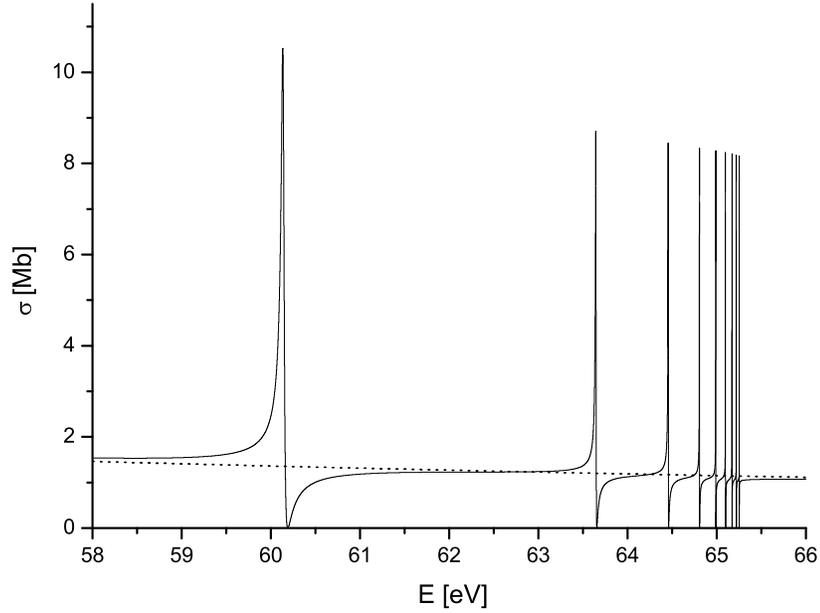}
\caption{Fano profiles for the $(sp,2n+)$ autoionizing series of Helium. The cross-sections are calculated following \cite{MORG}.
} \label{crossec}
\end{figure}

\subsection{Doppler broadening}

Due to the finite temperature $T$ of the gas, the frequency distribution of an ensemble of radiating atoms will experience a spread. Assuming the gas ensemble in thermal equilibrium, the Doppler shift is linked to the Maxwell-Boltzman probability distribution of velocities of the atoms. As a result, the Fano line in Eq. (\ref{sigsig2}) should be convolved with the Doppler-broadening profile $\mathcal{P}$ \cite{LODO}

\begin{eqnarray}
\mathcal{P}(E,E';T) = \frac{\exp\left[\frac{-M c^2 (E'-E)^2}{2 k T E^2}\right]}{\sqrt{\frac{2 \pi k T E^2}{M c^2}}}~,
\label{PE}
\end{eqnarray}
where $k$ is the Boltzmann constant, $T$ is the temperature, and $M$ is the molecular weight, yielding the Doppler-broadened cross-section

\begin{eqnarray}
\sigma_D(E) = \int d E' \sigma(E') \mathcal{P}(E,E';T)
\label{convosig}
\end{eqnarray}
From Eq. (\ref{PE}) we see that the distribution of frequencies turns out to be a Gaussian, with a peak at the resonance frequency and a FWHM given by

\begin{eqnarray}
\Gamma_D = 2 E_R \sqrt{\frac{2 (\mathrm{ln}2) kT}{M c^2}} \simeq 7 \cdot 10^{-7} E_R \sqrt{T[^\circ K]/M}~.
\label{doppler}
\end{eqnarray}
Assuming a temperature of $300$ K, the Doppler broadening is pressure independent and turns out to amount to about $0.4$ meV for He (molecular weight $M = 4$). For the resonances under study, $\Gamma_D$ is substantially smaller than the natural width $\Gamma \sim 4-10$ meV. It follows that the Fano profiles constitute a good approximation to the exact profile, which should be obtained by convolving the Fano natural line with the Gaussian profile due to Doppler shift. However, in order to quantitatively account for deviations of the peak efficiency of our monochromator\footnote{Doppler broadening is the only effect responsible for the deviation of the peak efficiency from $100 \%$. For example, collision broadening is very small. In fact we should compare collision times in the order of $100$ ns with the typical inverse ionization rate in the order of $100$ fs. As a result,  we expect an influence in the order of $0.0001 \%$, which can be neglected.} from $100\%$, in simulations we account for Doppler broadening effects according to Eq. (\ref{convosig}).

\section{\label{sec:qua} Amplitude and phase of the gas cell transmittance}

\begin{figure}[tb]
\includegraphics[width=1.0\textwidth]{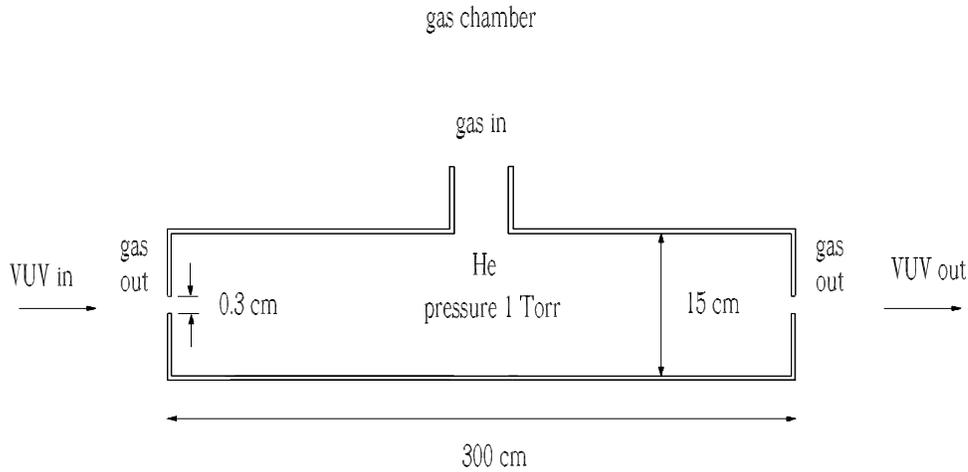}
\caption{Schematic of the gas cell} \label{gcell}
\end{figure}

\begin{figure}[tb]
\includegraphics[width=1.0\textwidth]{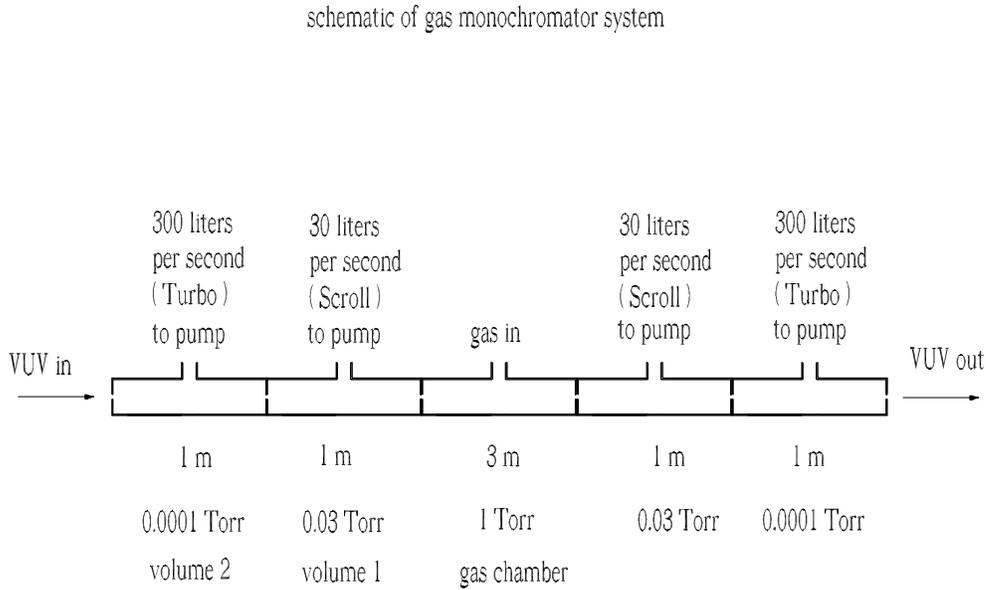}
\caption{Schematic of the differential pumping system} \label{pumps}
\end{figure}

\begin{figure}
\includegraphics[width=0.5\textwidth]{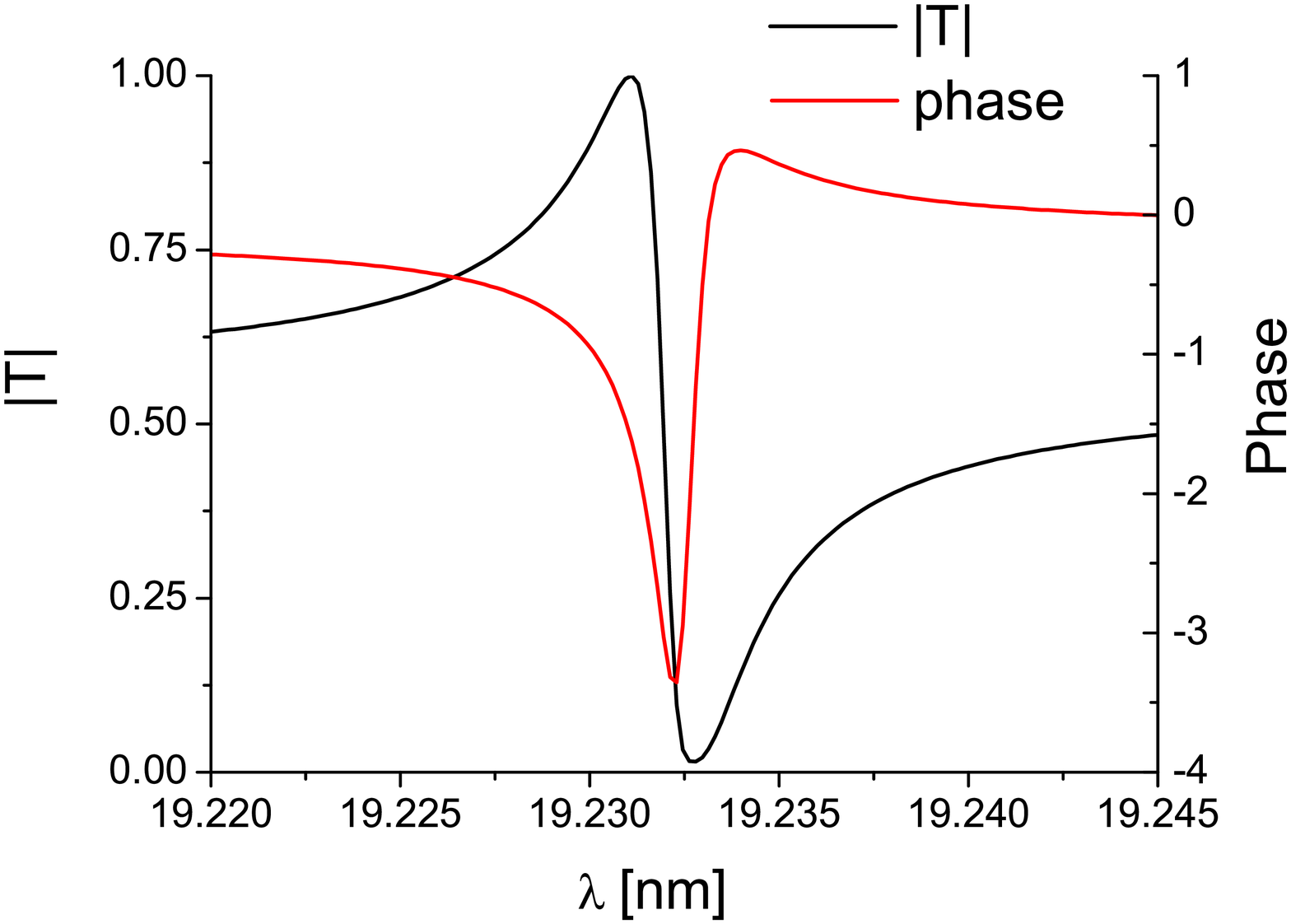}
\includegraphics[width=0.5\textwidth]{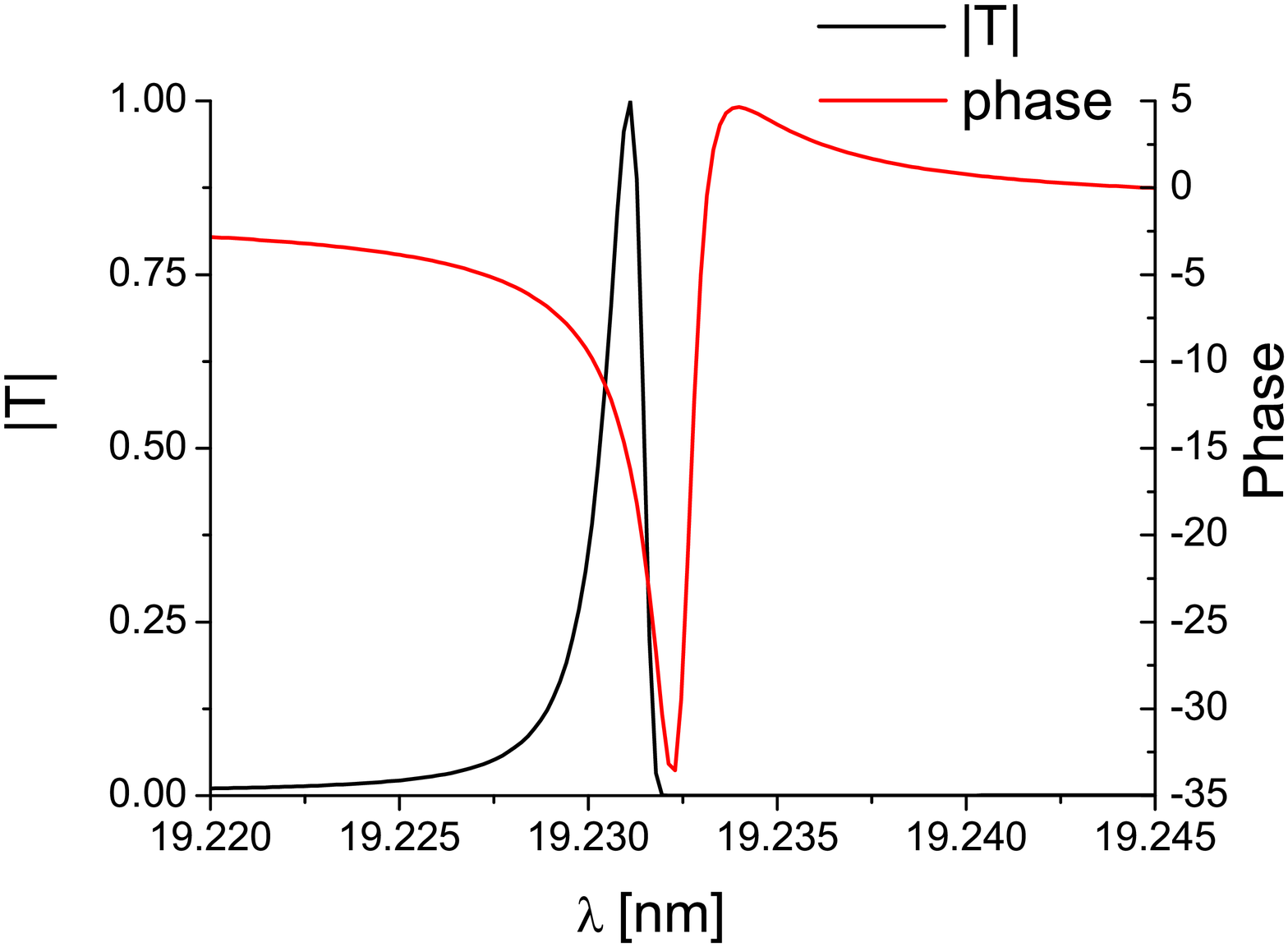}
\caption{Modulus and phase of the transmissivity of Helium around the $n=4$ line of the $(sp,2n+)~^1P_0$ Rydberg series. The modulus has been calculated according to Eq. (\ref{ModT}), while the phase is recovered with the help of the Kramers-Kroning relation according to Eq. (\ref{KKrel2}). The left plot refers to a column density $n_0 l = 10^{{18}} \mathrm{cm}^{-2}$. The right plot refers to a column density $n_0 l =  10^{19} \mathrm{cm}^{-2}$.}\label{transm}
\end{figure}

As discussed before, if a monochromatic electromagnetic pulse of intensity $I_0$ and frequency $\omega$ impinges on a cell of length $l$, filled with a gas with density $n_0$, the transmitted intensity obeys the Beer-Lambert law, Eq. (\ref{beerl}). As a result, the modulus of the transmissivity can be defined as

\begin{eqnarray}
|T| = \exp[-n_0 l \sigma/2]~,
\label{ModT}
\end{eqnarray}
where $\sigma(\omega)$ follows a Fano profile convolved with the Doppler-broadening profile, as has been discussed in the previous Section. In this article we will assume that the gas used in the monochromator is Helium, and that the energy of the FEL photons is within the range between $60$ eV and $65$ eV. These requirements ensure that the cross-section has a zero for $\mathcal{E} = - q$, meaning that the correlation parameter $\rho^2 = 1$. In order to obtain a pass-band filter, one needs to increase the gas pressure so to absorb radiation outside a small bandwidth from the point where $\mathcal{E} = - q$. The pressure in the gas cell is limited by capabilities of the differential pumping system, and cannot exceed a few Torr in the present design. Fig. \ref{gcell} and Fig. \ref{pumps} show the schematic of the gas cell, and its integration in the differential pumping
system.  The design presented here is based on \cite{RYU1}-\cite{RYU5}, which were used in the design of the LCLS gas attenuator. The length of the gas cell is between $3$ m and $5$ m, Fig. \ref{gcell}, and cannot be easily increased due to space limitations. Even with these limitations on length and pressure, the gas cell can provide a very large attenuation coefficient. For example, at a cell length of $3$ m and a pressure of $1$ Torr the column density is  $n_0 l \sim 10^{19} \mathrm{cm}^{-2}$, and the spectrum of the incoming radiation is attenuated by the background cross section of about $40$ dB.

Let us consider the third ($n=4$) resonant line of the $(sp,2n+)~^1P_0$ Rydberg series for Helium, and let us calculate its cross-section with the help of\footnote{As already said, in actual simulations we will also account for Doppler broadening effects. For simplicity we do not account for them here.} Eq. (\ref{sigsig2}) and Eq. (\ref{sigb}). The modulus of the transmissivity can be found with the help of Eq. (\ref{ModT}). In order to exemplify the effects of the pressure increase, we select two values for the column-density: $n_0 l =  10^{18} \mathrm{cm}^{-2}$, and $n_0 l =  10^{19} \mathrm{cm}^{-2}$. Since $|T|$ is known, it is possible to use the Kramers-Kroning relations to recover the phase.  In fact one can write

\begin{eqnarray}
\mathrm{ln}[T(\omega)] = \mathrm{ln}[|T(\omega)|] + i
\Phi(\omega) = - nl \sigma/2 + i
\Phi(\omega)~.\label{ln}
\end{eqnarray}
Note that $T^*(\omega)=T(-\omega)$ implies that $|T(\omega)|=|T(-\omega)|$ and that $\Phi(\omega) = - \Phi(-\omega)$. Therefore, using Eq. (\ref{ln}) one also has that $\mathrm{ln}[T(\omega)]^*=\mathrm{ln}[T(-\omega)]$. Then, application of Titchmarsh theorem shows that the analyticity of $\mathrm{ln}[|T(\Omega)|]$ on the upper complex $\Omega$-plane implies that

\begin{eqnarray}
\Phi(\omega)=-\frac{2\omega}{\pi}\mathcal{P} \int_0^{\infty}
\frac{\mathrm{ln}[T(\omega')] }{\omega'^2-\omega^2} d\omega'~,
\label{KKrel2}
\end{eqnarray}
A direct use of Eq. (\ref{KKrel2}), with $|T|$ given as in Eq. (\ref{ModT}), yields back the phase $\Phi(\omega)$. By this, one tacitly assumes that $\mathrm{ln}[T(\Omega)]$ is analytical on the upper complex $\Omega$-plane. This fact, however, is immediately granted by the fact that $\sigma$ is proportional to the imaginary part of the refractive index of the medium, and it is well known that the refractive index must obey the Kramers-Kroning relation Eq. (\ref{KKrel2}). We therefore used Eq. (\ref{KKrel2}) in order to recover the phase of the transmittance. More specifically, we took advantage of a publicly available Matlab script \cite{LUC2} which serves exactly to that end. The final result in terms of modulus and phase of the transmissivity $T$ is shown in Fig. \ref{transm} for the case $n_0 l =  10^{18} \mathrm{cm}^{-2}$, left plot\footnote{If the reader will compare this plot with Fig. 5 of \cite{SOFT}, the reader will notice differences in $|T|$. This is due to a misprint in Fig. 5 of \cite{SOFT}, where $|T|^2$ and not $|T|$ is actually plotted.} and $n_0 l =  10^{19} \mathrm{cm}^{-2}$, right plot. One can see that an increase in the gas density of a factor $10$ transform the filter in Fig. \ref{transm} left, into the few meV bandpass filter in Fig. \ref{transm} right.

The example studied here for the $n=4$ resonance case corresponds to a pass-band filter with a $2.1$ meV FWHM-bandwidth. As said above, such bandwidth is narrower than that of a typical FEL mode, and this enables the delivery of single-mode VUV laser beams. Note that the knowledge of modulus and phase of the transmittance allows for the characterization of the single-mode output both in modulus and phase. This is an obvious advantage for scientists wishing to model the interaction between single-mode VUV pulses and matter.

\section{\label{sec:cin} Simulations of gas monochromator operation at FLASH}

\begin{figure}[tb]
\includegraphics[width=0.5\textwidth]{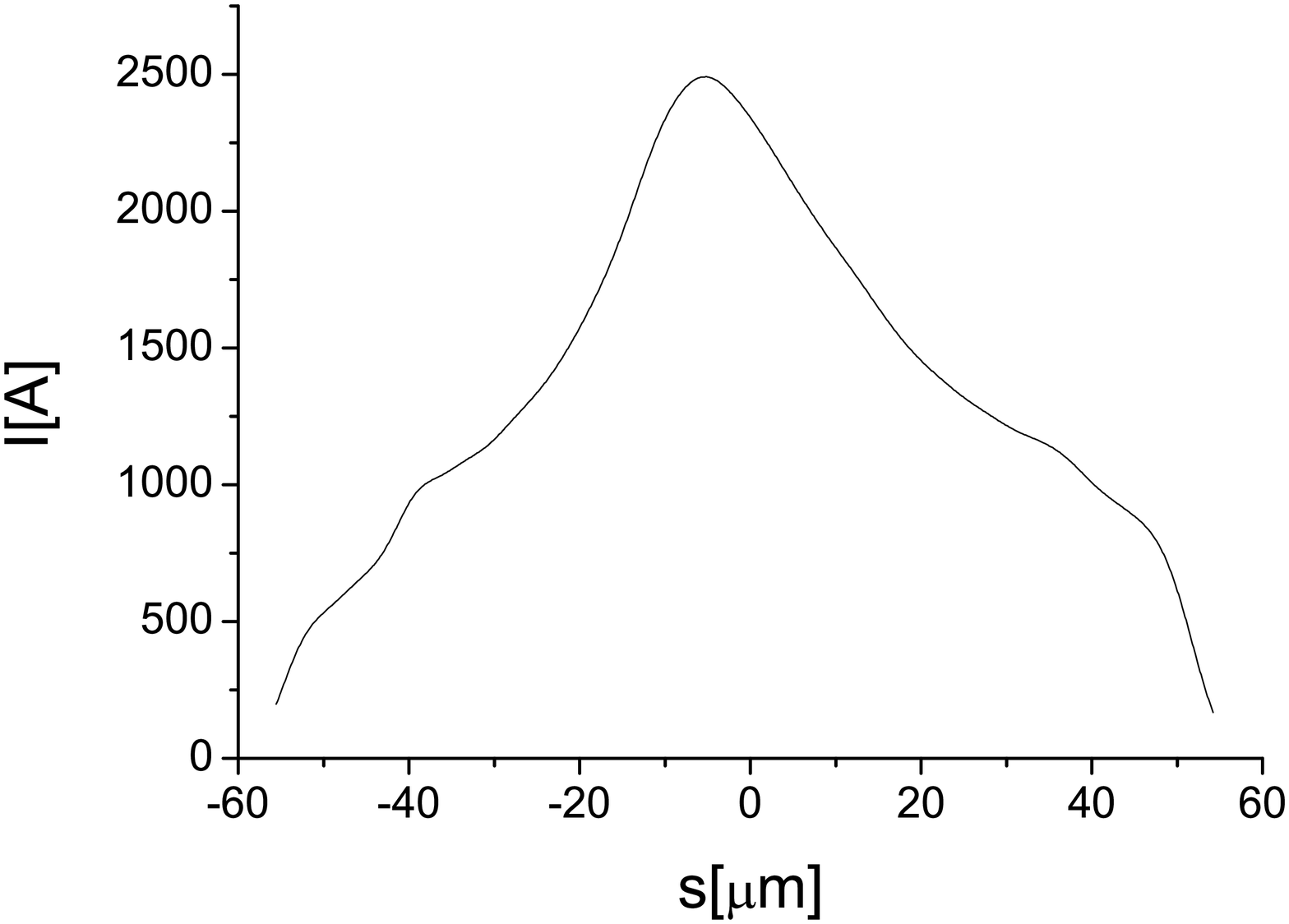}
\includegraphics[width=0.5\textwidth]{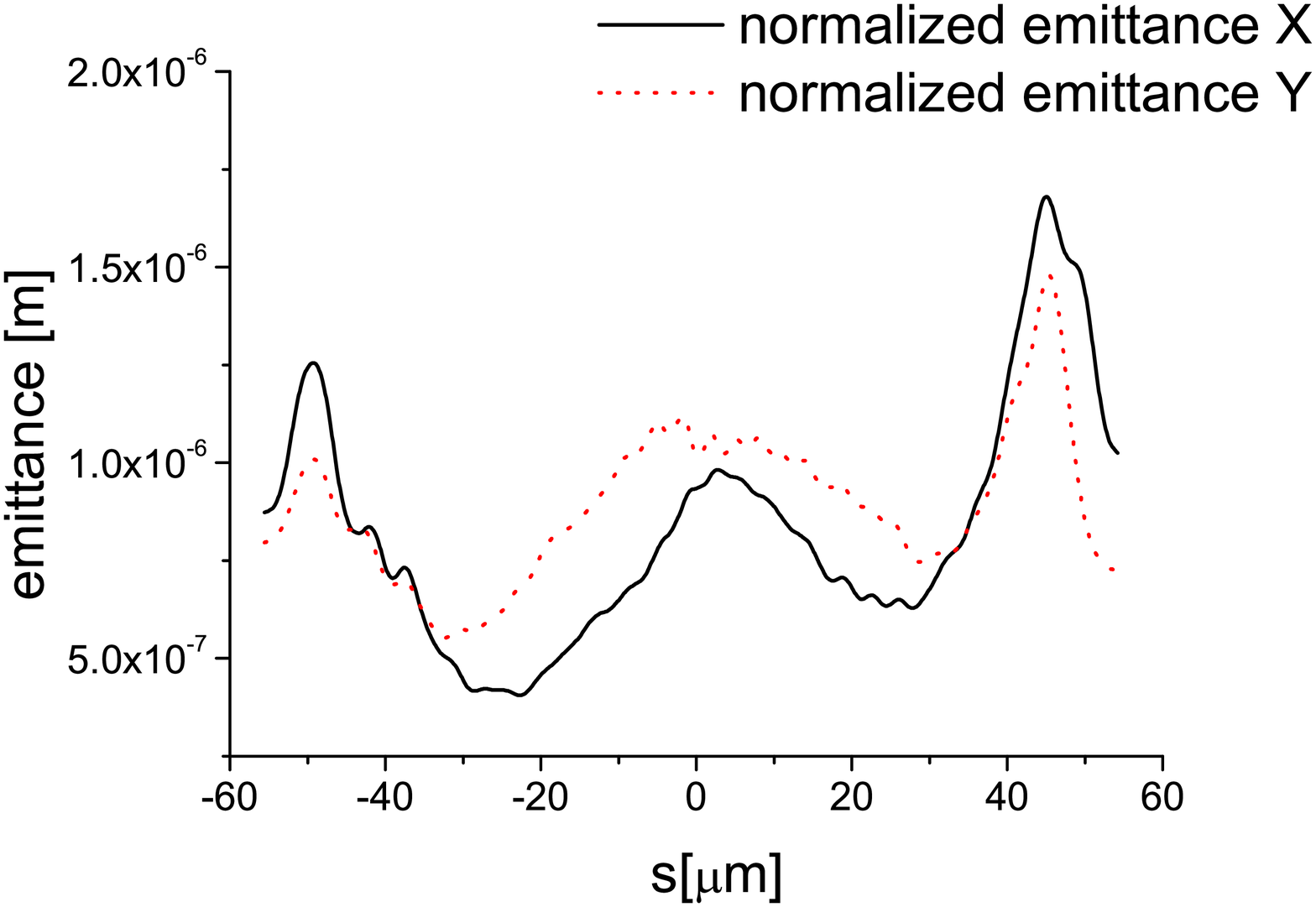}
\includegraphics[width=0.5\textwidth]{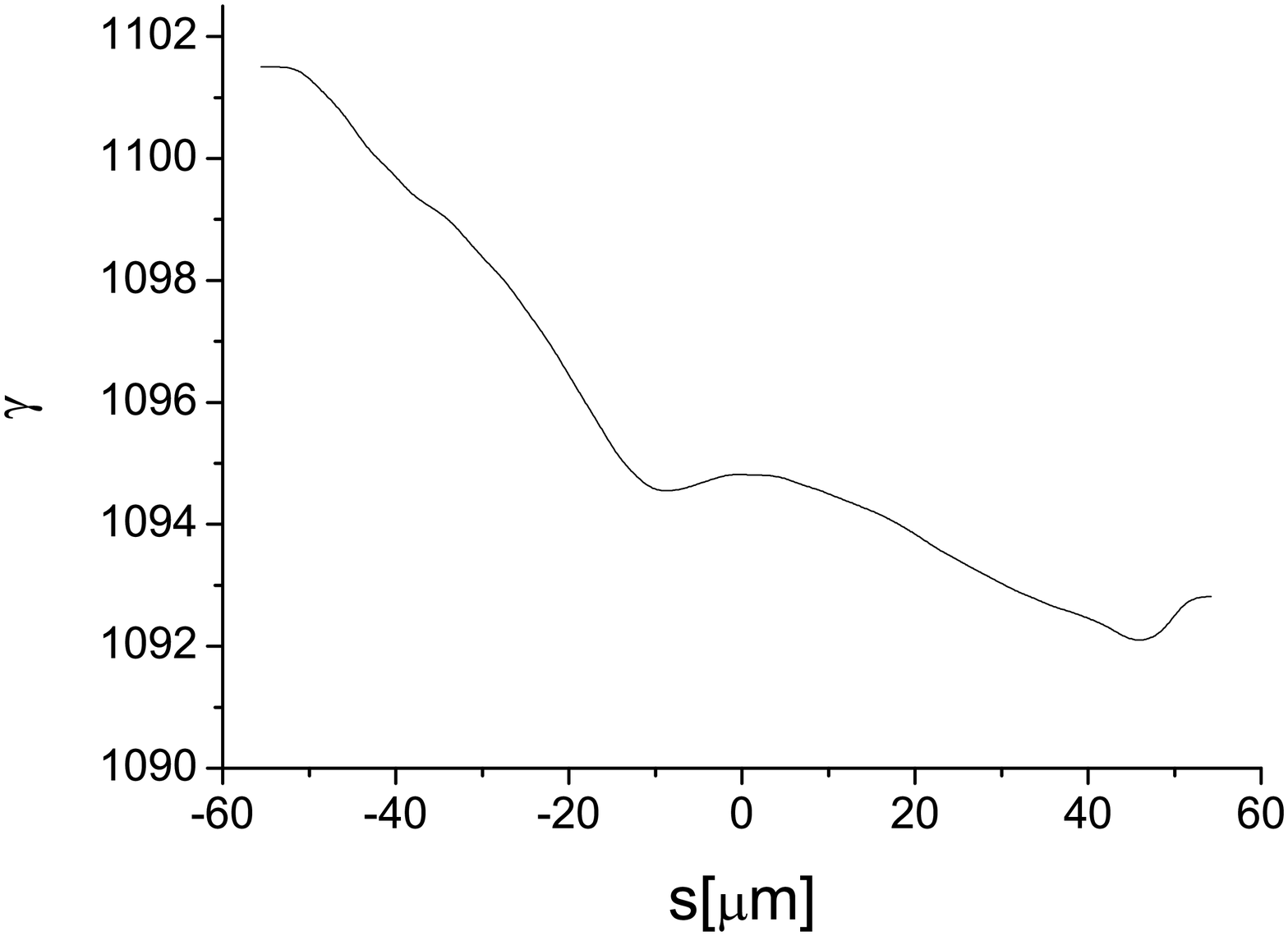}
\includegraphics[width=0.5\textwidth]{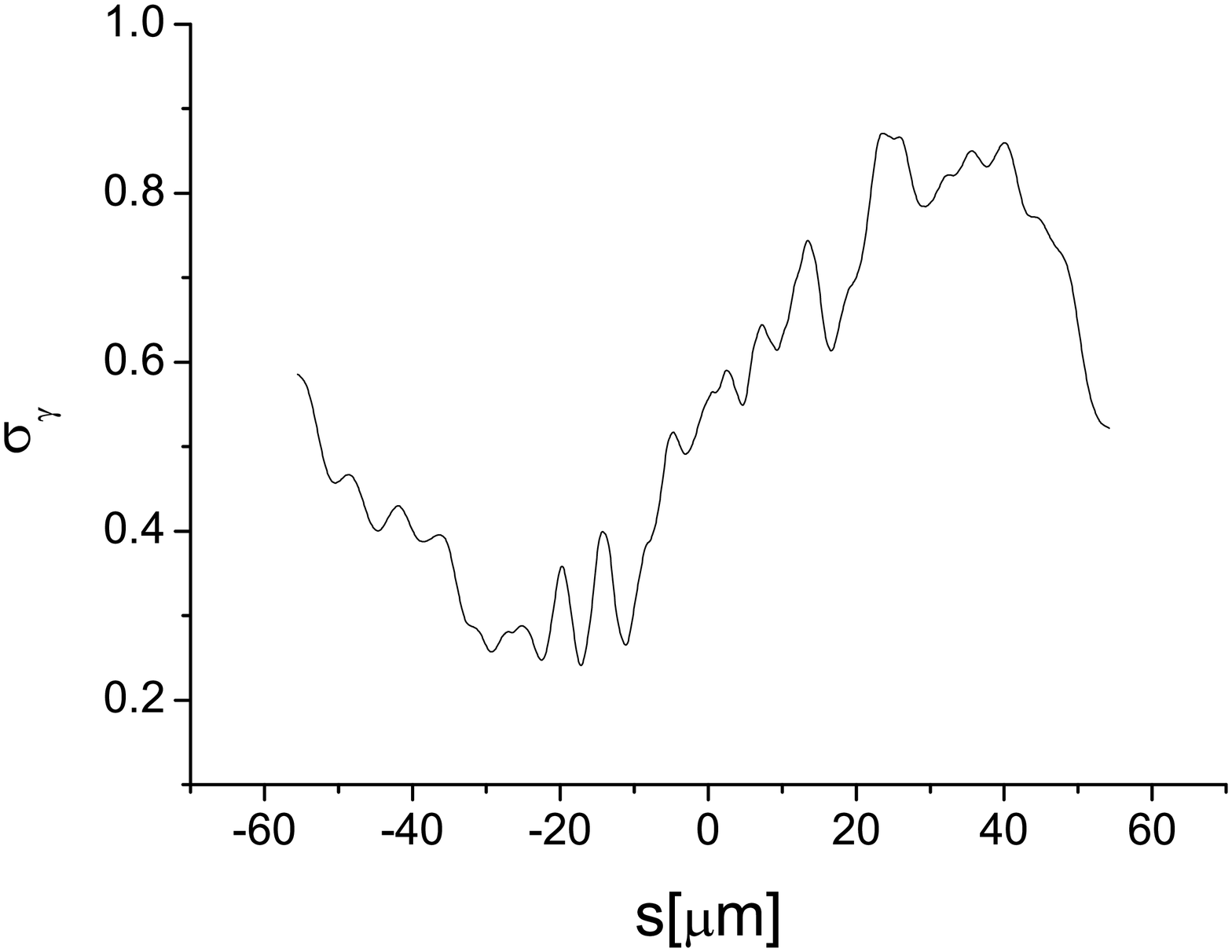}
\caption{Electron beam characteristics at the entrance of the setup at the FLASH. (upper left) Current profile. (upper right) horizontal and vertical geometrical emittance. (lower left) energy profile. (lower right) rms energy spread profile.} \label{ebeam}
\end{figure}

\begin{figure}[tb]
\includegraphics[width=0.5\textwidth]{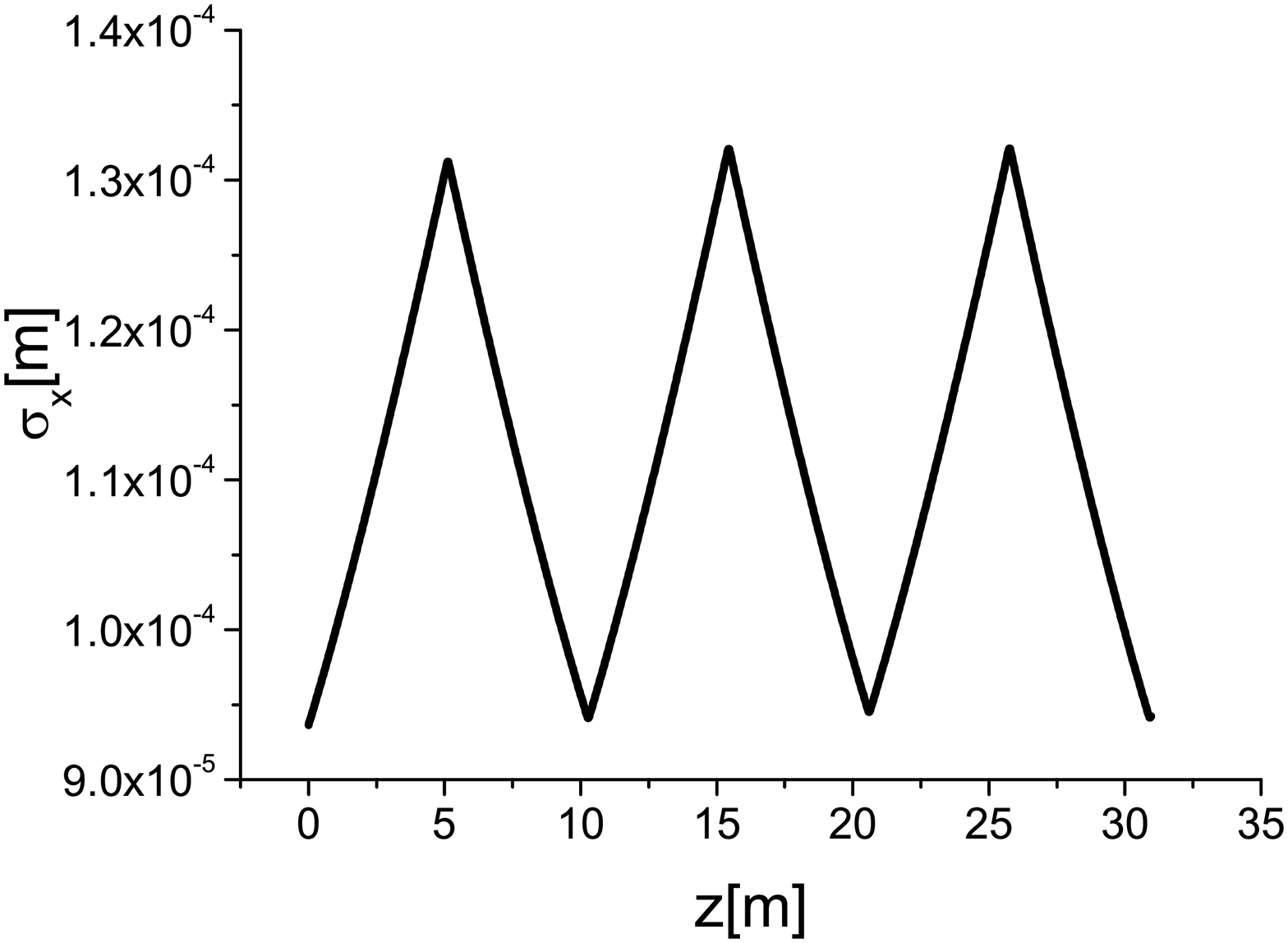}
\includegraphics[width=0.5\textwidth]{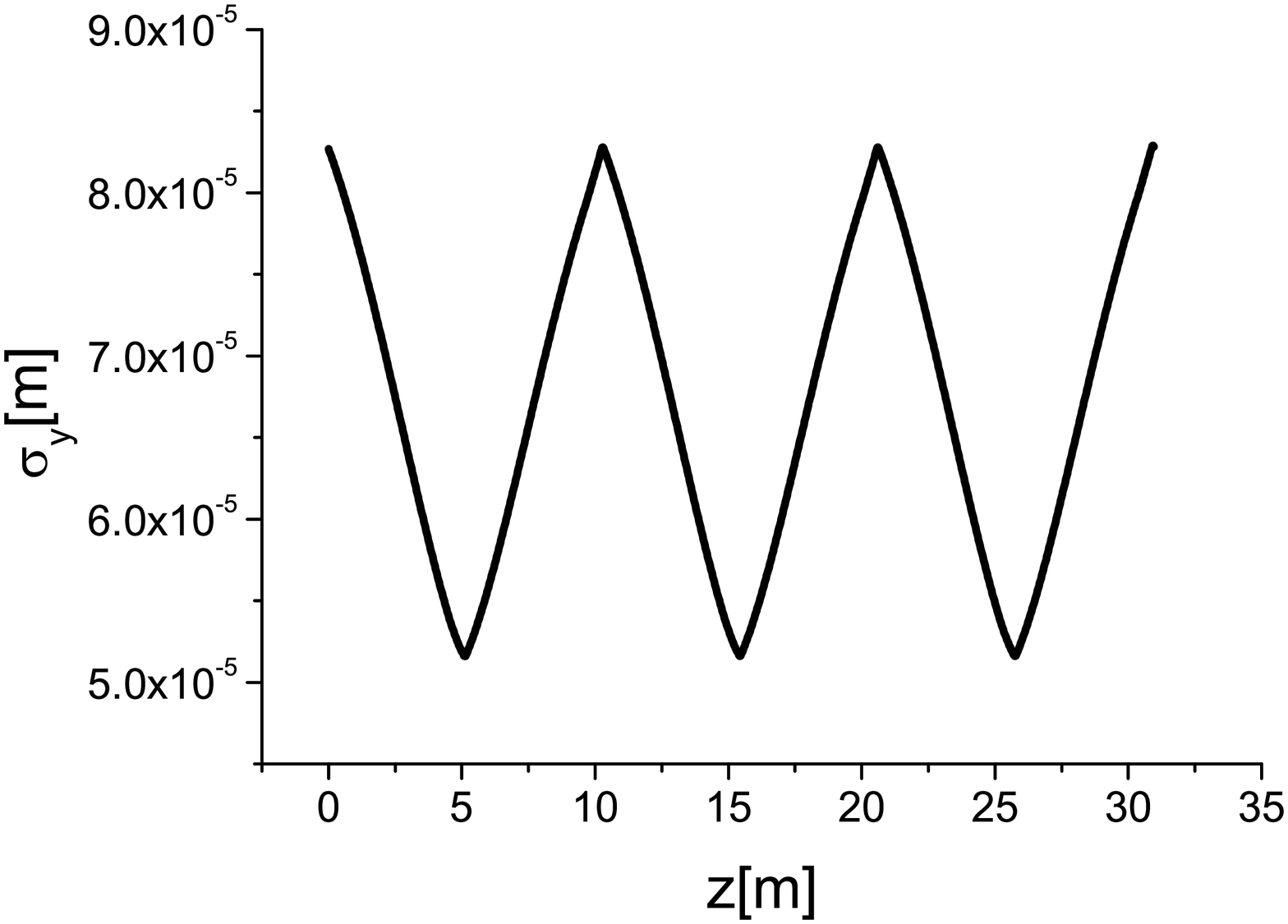}
\caption{Electron beam horizontal (left plot) and vertical (right plot) rms sizes as a function of the distance inside the undulator.} \label{size}
\end{figure}

\begin{table}
\caption{Parameters for the nominal pulse mode of operation used in
this paper.}

\begin{small}\begin{tabular}{ l c c}
\hline & ~ Units &  ~ \\ \hline
Undulator period      & mm                  & 27.3     \\
K parameter (rms)      &                    & 0.89  \\
Wavelength  (fundamental)               & nm                  & 20   \\
Charge                & nC                  & 0.5 \\
Electron beam energy & MeV                  & 560 \\
\hline
\end{tabular}\end{small}
\label{tt2}
\end{table}
In this section we present a feasibility study of our scheme. We demonstrate the potential of our technique using as an example the parameters for the FLASH facility in the VUV region, although our setup can be exploited by other FEL and synchrotron radiation facilities in the same photon energy range. The main parameters  for the FLASH facility operating around are listed in Table \ref{tt2}.  Full 3D simulations have been performed in order to confirm the scheme feasibility. The simulations are performed with the code GENESIS 1.3 \cite{GENE}, which uses as input the beam parameters obtained in start to end simulations \cite{echo}. The beam distributions are shown in Fig. \ref{ebeam} showing the current profile, the horizontal and vertical normalized emittances, the energy profile and the rms energy spread profile. The focusing system along the FLASH undulator consists of doublets between each undulator segment. The three possible focusing solutions are a doublet structure, a FOFO structure and a FODO structure \cite{FOCU}. In our simulations we implemented the FODO focusing type. The average betatron function used were about $10$ m, resulting in the beam rms size shown in Fig. \ref{size} as a function of the distance inside the undulator.

In the following we will consider different resonances, and different possible settings of FLASH, in order to analyze several configurations of the gas monochromator setup.

\subsection{First resonance at $60.14$ eV}

\begin{figure}[tb]
\includegraphics[width=0.5\textwidth]{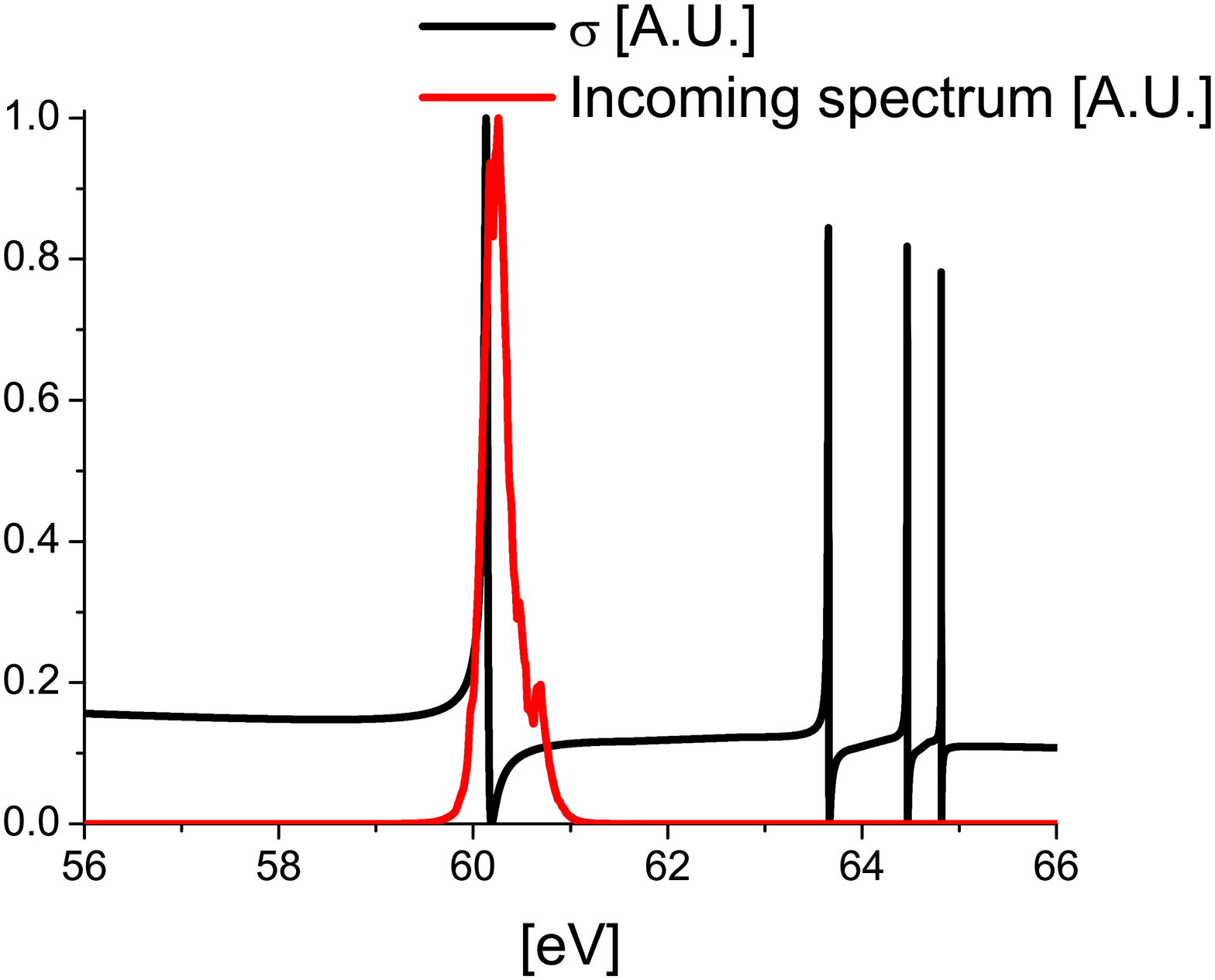}
\includegraphics[width=0.5\textwidth]{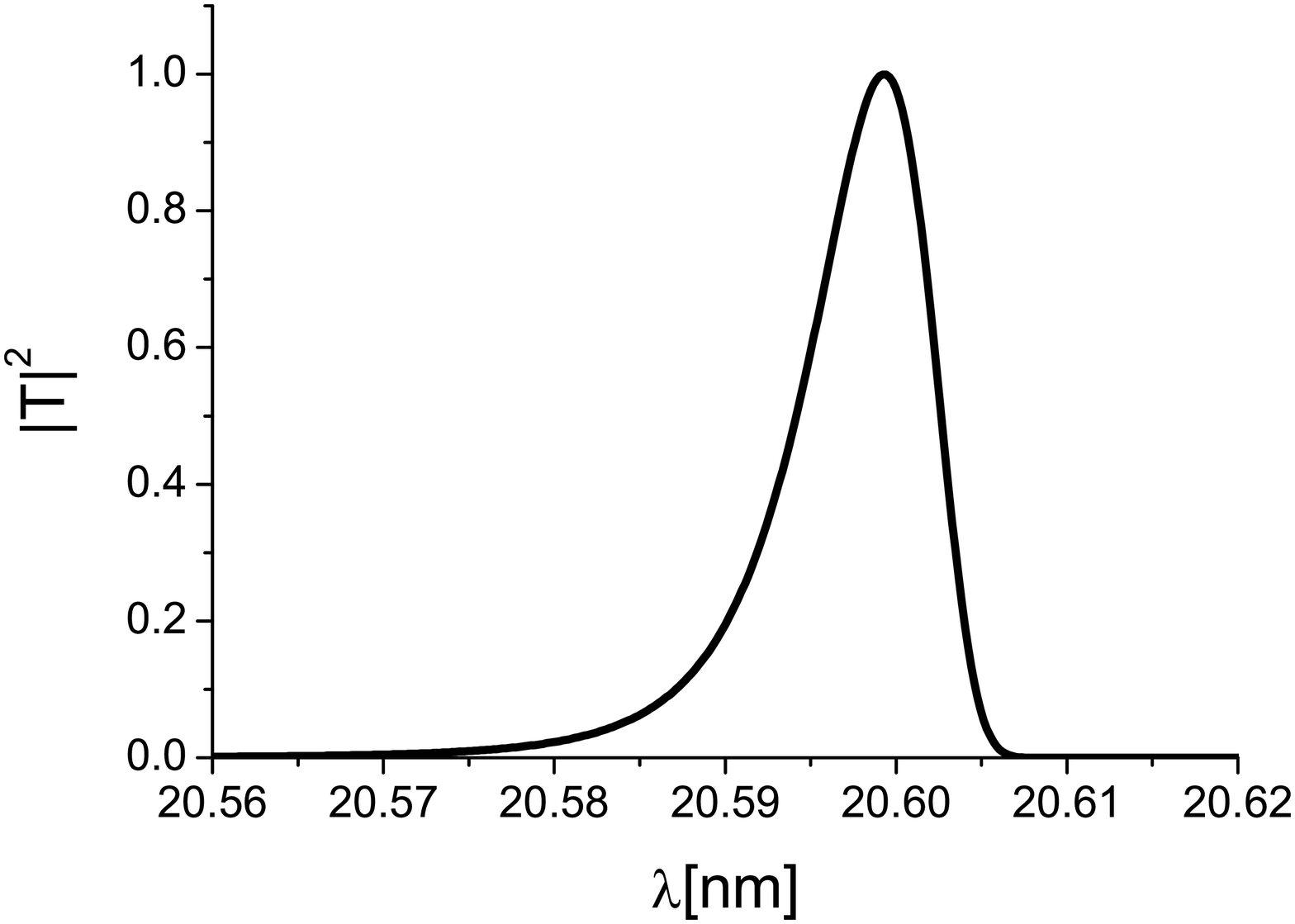}
\caption{Left plot: the SASE spectrum at saturation overlaps the first resonance. Right plot: the filter bandwidth for a column density of $n_0 l = 10^{19} \mathrm{cm}^{-2}$.} \label{Tsig}
\end{figure}

We begin considering the first autoionization profile $n=2$ for the $(sp,2n+)$ autoionizing series of Helium. FLASH is tuned at saturation, that is we consider three active undulator modules corresponding to a magnetic length of $15$ m, so that the SASE average bandwidth overlaps the first resonance, as shown in Fig. \ref{Tsig} left. The right plot in Fig. \ref{Tsig} shows, instead, the bandpass-filter profile in terms of $|T|^2$ for a column density of $n_0 l = 10^{19} cm^{-2}$.

As one can see, in this case the filter bandwidth is around $23$ meV FWHM, which is narrower, but in the same order of magnitude of the bandwidth of a single SASE spike. As a result, we should expect a dependency of the output spectra on the input intensity, which changes from shot to shot. The envelope of the output spectra will follow, however, the distribution in Fig. \ref{Tsig}, right.

\begin{figure}[tb]
\includegraphics[width=0.5\textwidth]{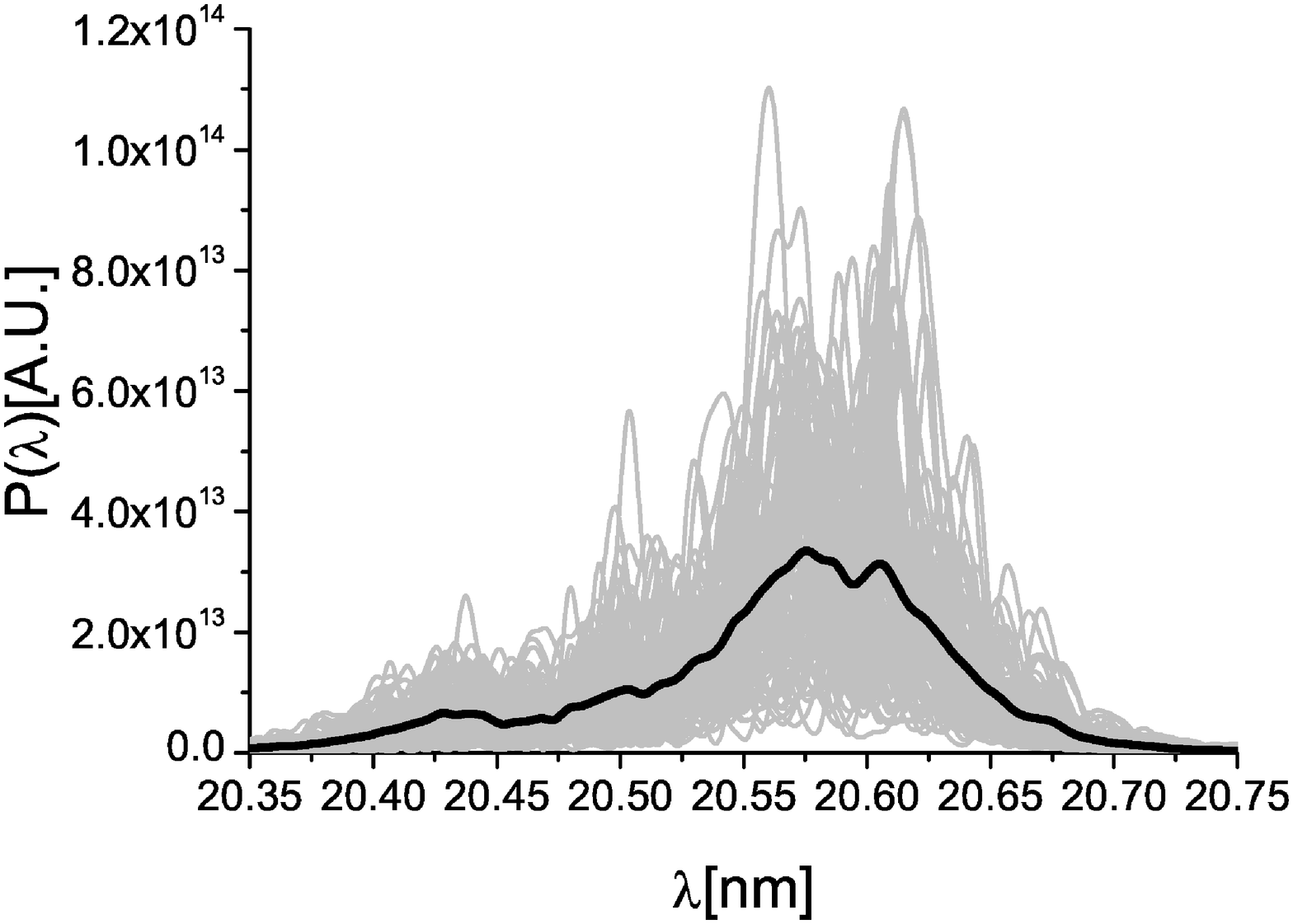}
\includegraphics[width=0.5\textwidth]{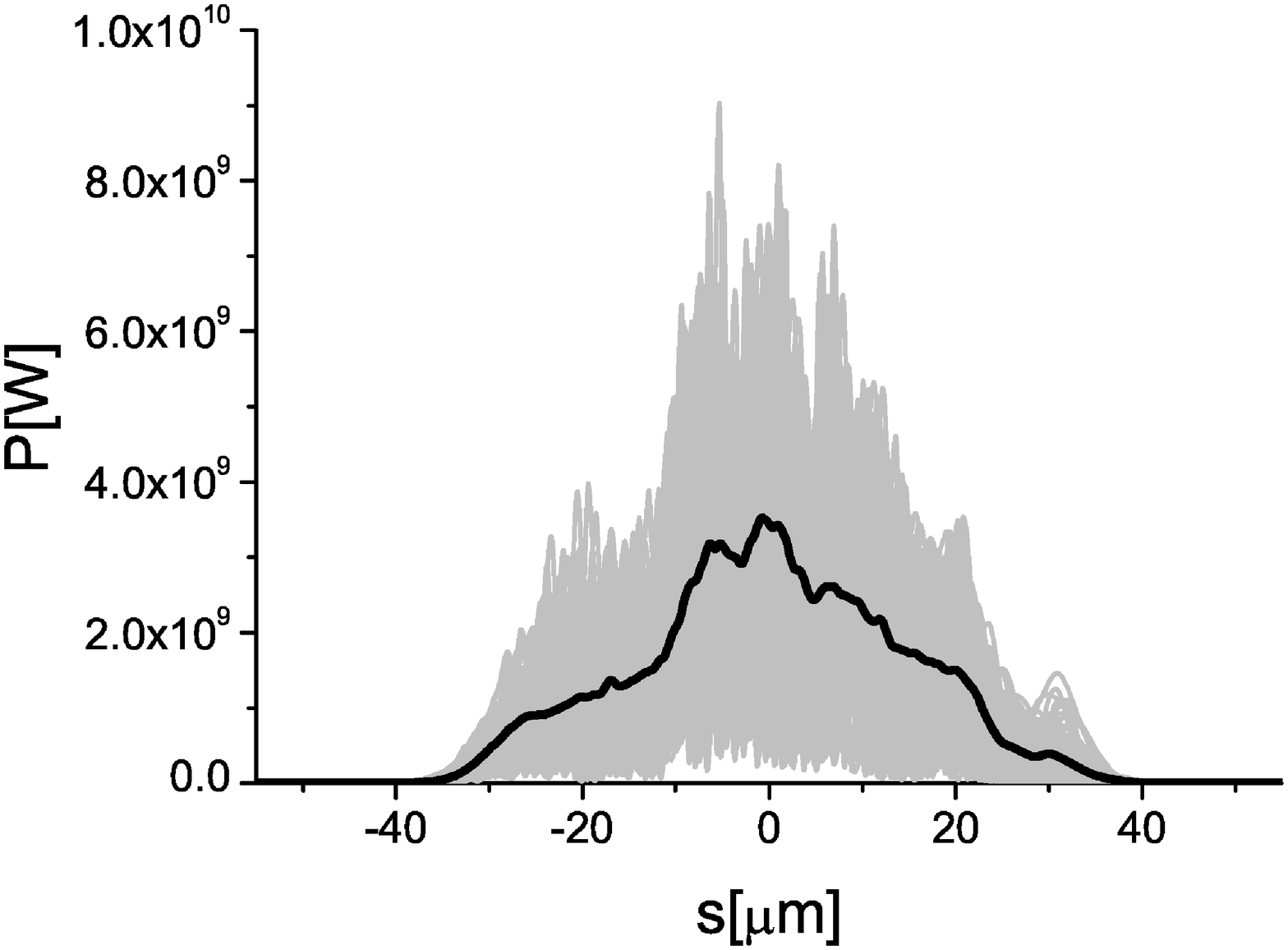}
\caption{First resonance mode of operation. Pulse spectrum (left plot) and power (right plot) before the gas cell. Grey lines refer to single shot realizations, the black line refers to an average over one hundred realizations.} \label{IN}
\end{figure}
The input spectra and power are shown respectively in the left and right plots of Fig. \ref{IN}.

\begin{figure}[tb]
\includegraphics[width=0.5\textwidth]{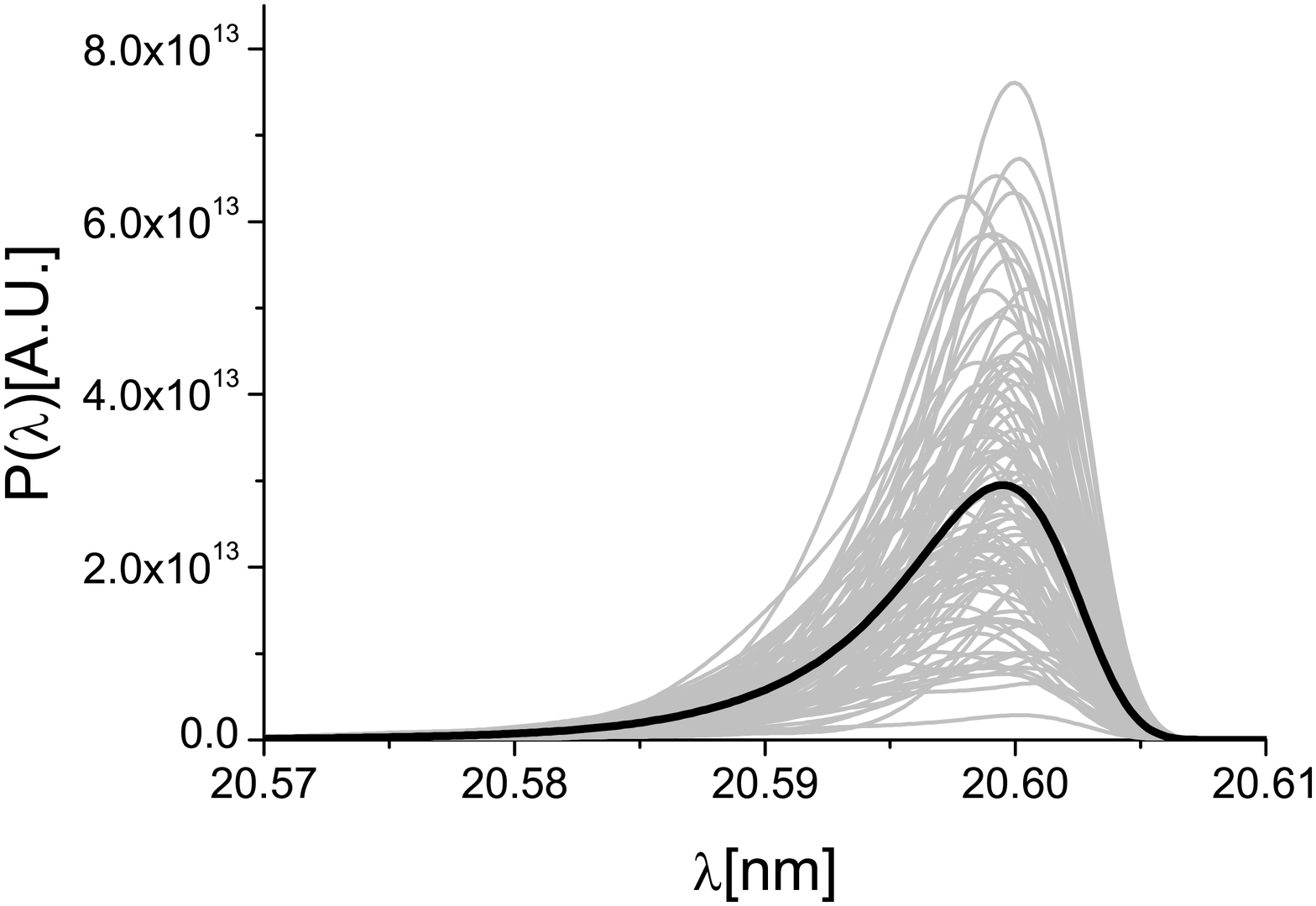}
\includegraphics[width=0.5\textwidth]{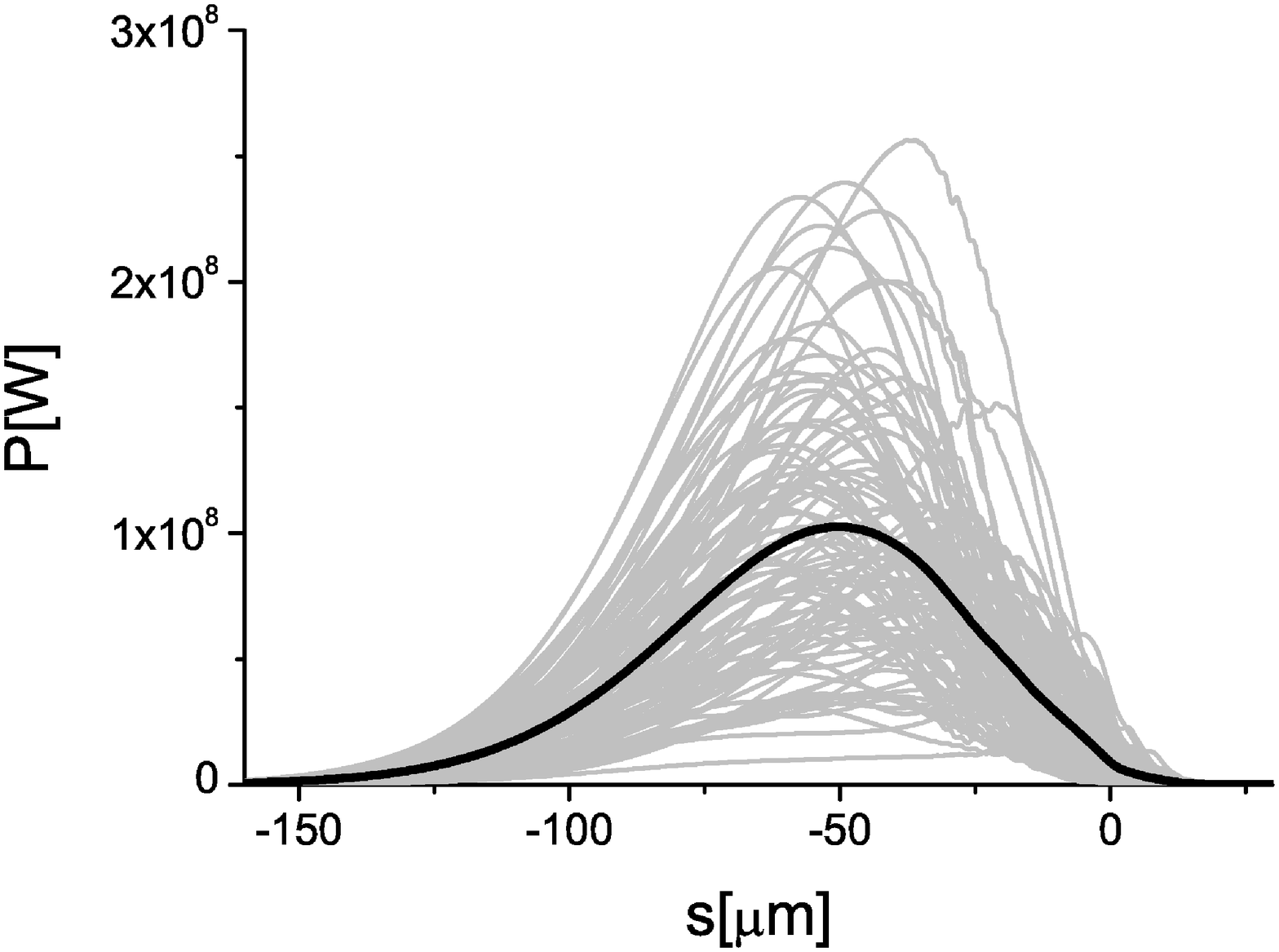}
\caption{First resonance mode of operation. Pulse spectrum (left plot) and power (right plot) after the gas cell. Grey lines refer to single shot realizations, the black line refers to an average over one hundred realizations.} \label{OUT}
\end{figure}
The effect of the filter is shown in Fig. \ref{OUT}, where we plot the output spectra (left plot) and powers (right plot). The shot-to-shot fluctuations in the profiles of both intensity and spectrum are a consequence of the fact that the filter bandwidth is of the same order of a single spike profile.

It should be noted that the theoretical calculations used to calculate the overall transmission function of the gas cell are based on the validity of the perturbation theory approach. In our case the transverse size of the photon beam at the cell is about $1$ mm (thus fitting into the orifices aperture of $3$ mm diameter, Fig. \ref{gcell}). The energy in the VUV pulse can be evaluated from Fig. \ref{IN}, and is about $0.3$ mJ. At VUV wavelengths, this means that a pulse contains about $10^{14}$ photons. The photon density is thus $10^{16} \mathrm{cm}^{-2}$.   It should be noted that the resonance line is about two orders of magnitude narrower($10^{-4}$) compared with the SASE bandwidth ($5 \cdot 10^{-3}$), while the maximum cross section reaches about 10 Mb. As a result, ionization takes place mainly due to the background cross section, of about 1.5 Mb. Recalling that 1 Mb is $10^{18} \mathrm{cm}^2$ one can confirm the validity of the perturbation theory approach used to calculate the overall transmission function of the gas cell.

This reasoning is valid for a single radiation pulse. Limitations for time diagram of facility operation can be estimated with the help of \cite{RYYY}, where space-charge effects are studied, and an estimation of ambipolar diffusion is presented. Based on this study we can estimate that the single bunch mode operation with a 10 Hz repetition rate at FLASH will not be problematic, while for other facilities it should be possible to stretch this range up to about $100$ Hz. However, considering, for example, the FLASH multi bunch mode of operation ($1$ MHz rep rate within $1$ ms macro pulse, with $10$ Hz  macro pulse repetition rate) further studies on the diffusion process in the gas cell are needed, which are left for future work.

\subsection{Second resonance at $63.655$ eV}

\begin{figure}[tb]
\includegraphics[width=0.5\textwidth]{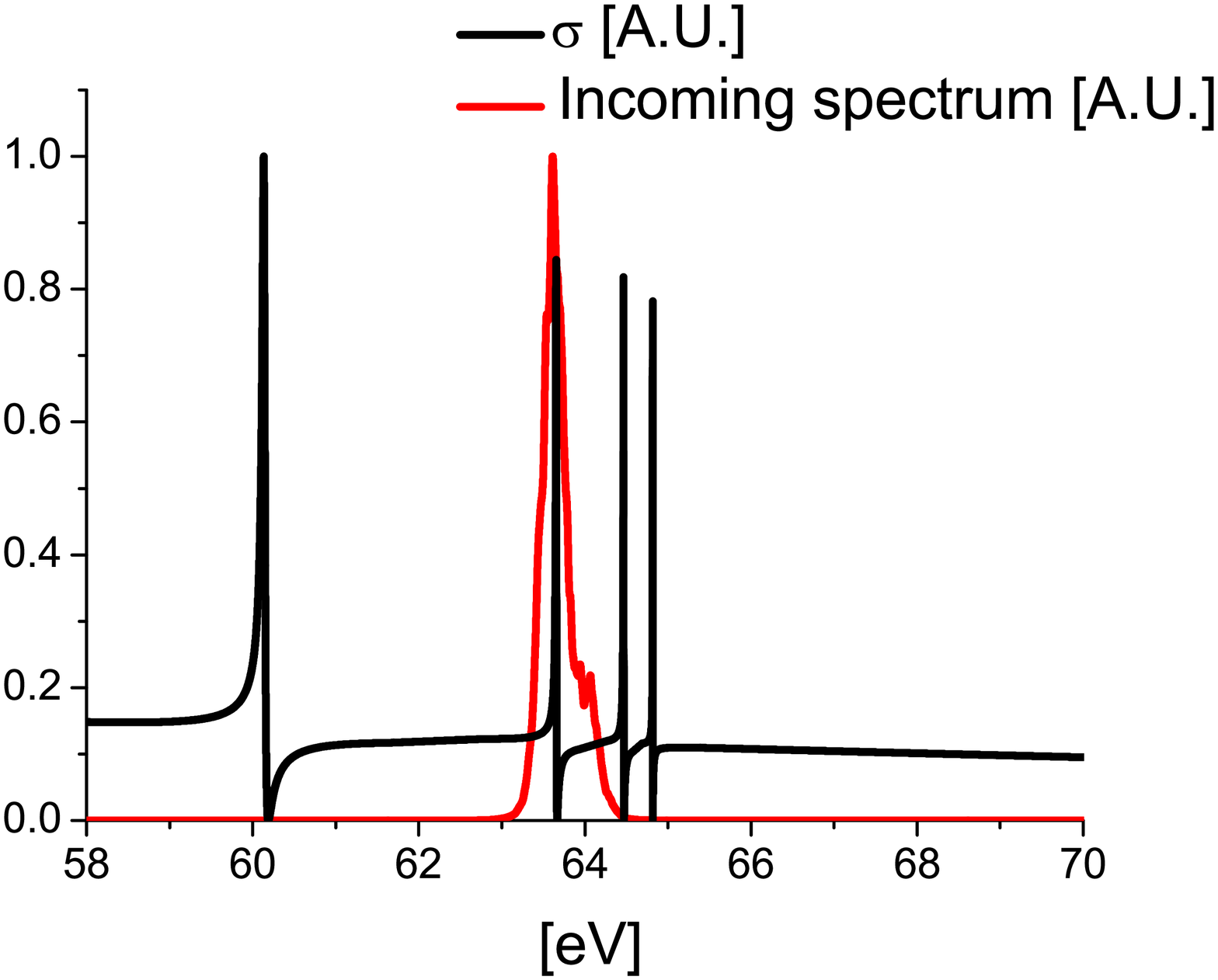}
\includegraphics[width=0.5\textwidth]{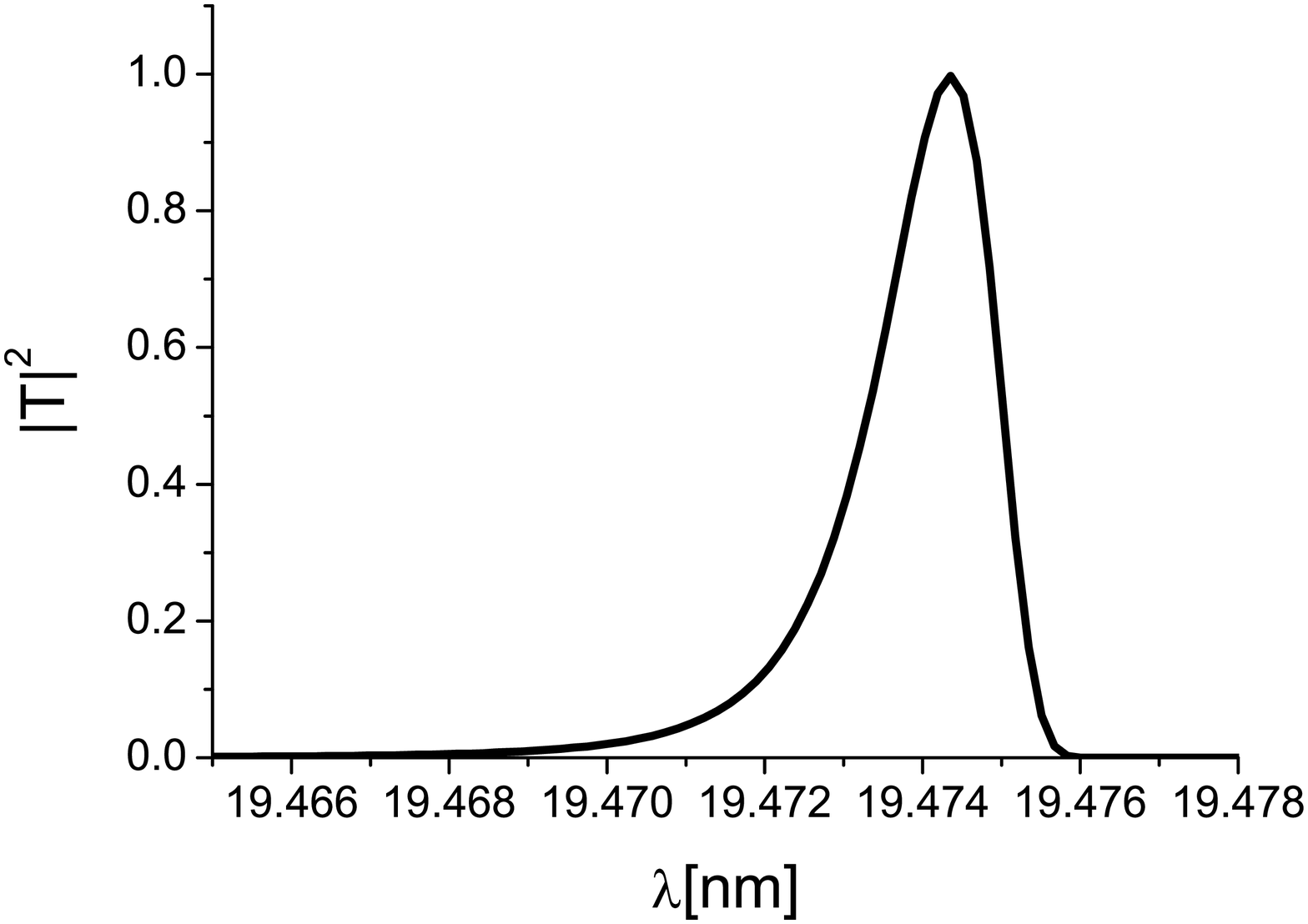}
\caption{Left plot: the SASE spectrum at saturation overlaps the second resonance. Right plot: the filter bandwidth for a column density of $n_0 l = 10^{19} \mathrm{cm}^{-2}$.} \label{Tsig2}
\end{figure}

We now consider the autoionization profile $n=3$ for the $(sp,2n+)$ autoionizing series of Helium. FLASH is tuned at saturation, so that the SASE average bandwidth overlaps the second resonance, as shown in Fig. \ref{Tsig2} left. The right plot in Fig. \ref{Tsig2} shows, instead, the bandwidth of the filter in terms of $|T|^2$ for a column density of $n_0 l = 10^{19} \mathrm{cm}^{-2}$.

\begin{figure}[tb]
\includegraphics[width=0.5\textwidth]{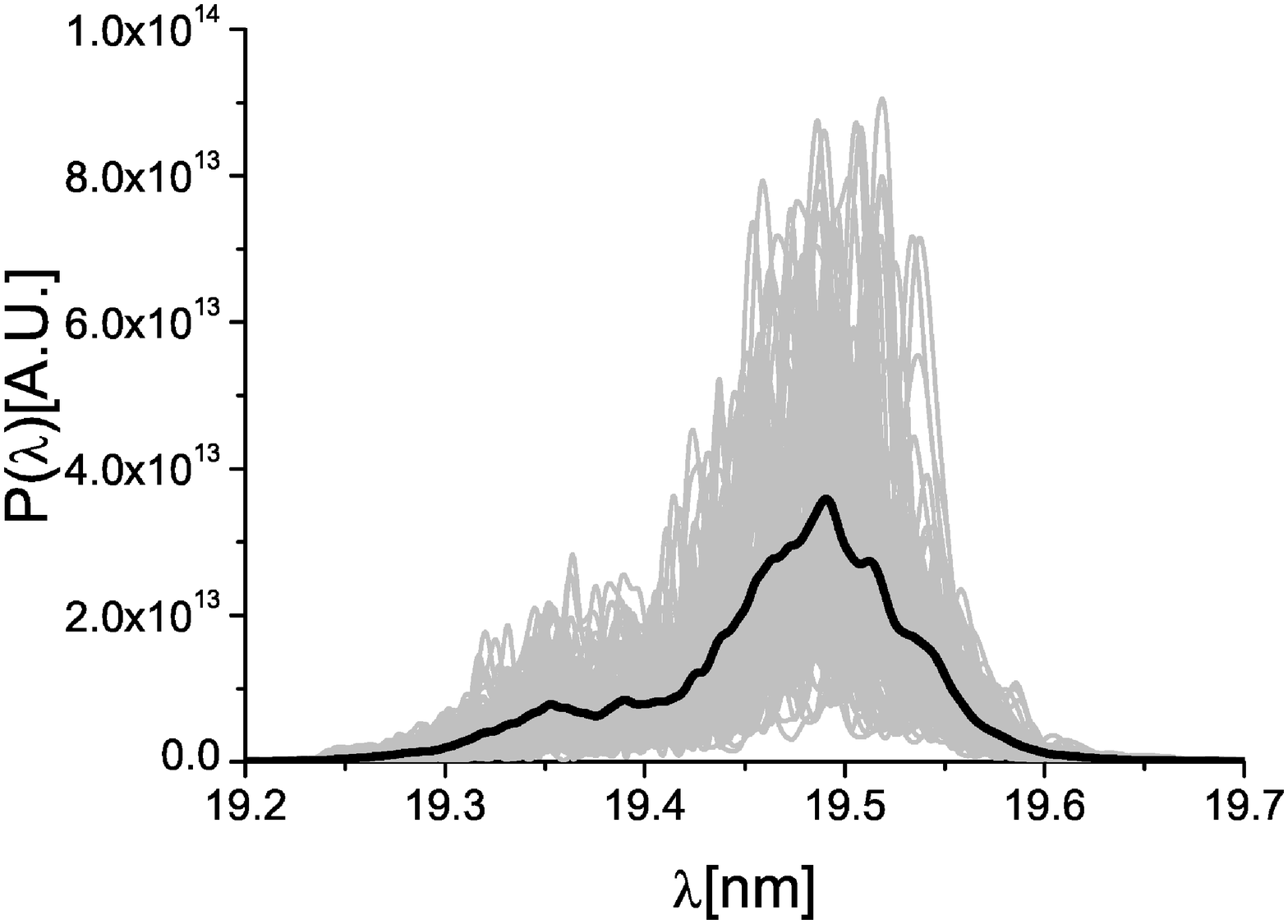}
\includegraphics[width=0.5\textwidth]{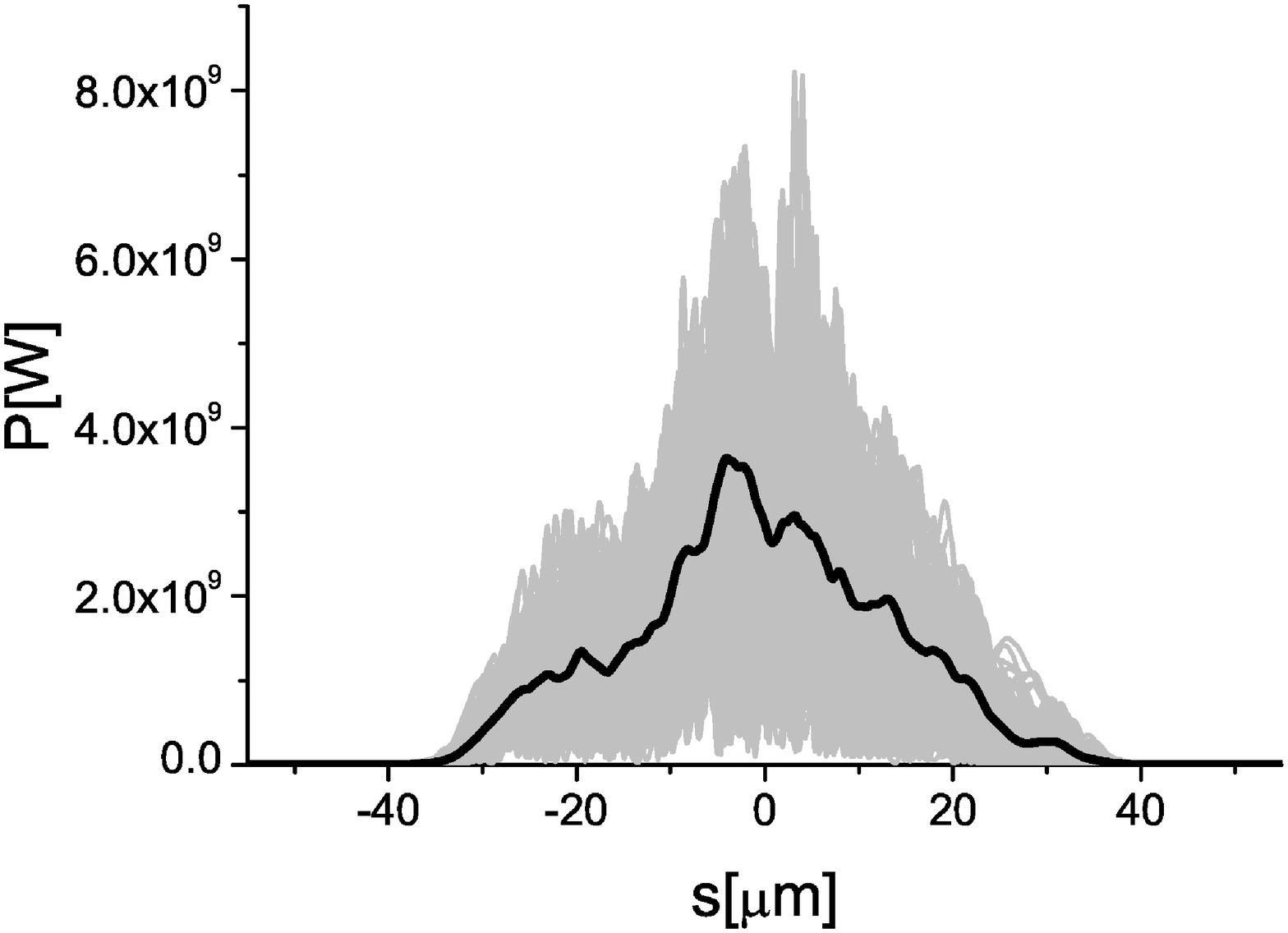}
\caption{Second resonance mode of operation. Pulse spectrum (left plot) and power (right plot) before the gas cell. Grey lines refer to single shot realizations, the black line refers to an average over one hundred realizations.} \label{IN2}
\end{figure}
The input spectra and power are shown respectively in the left and right plots of Fig. \ref{IN2}.

\begin{figure}[tb]
\includegraphics[width=0.5\textwidth]{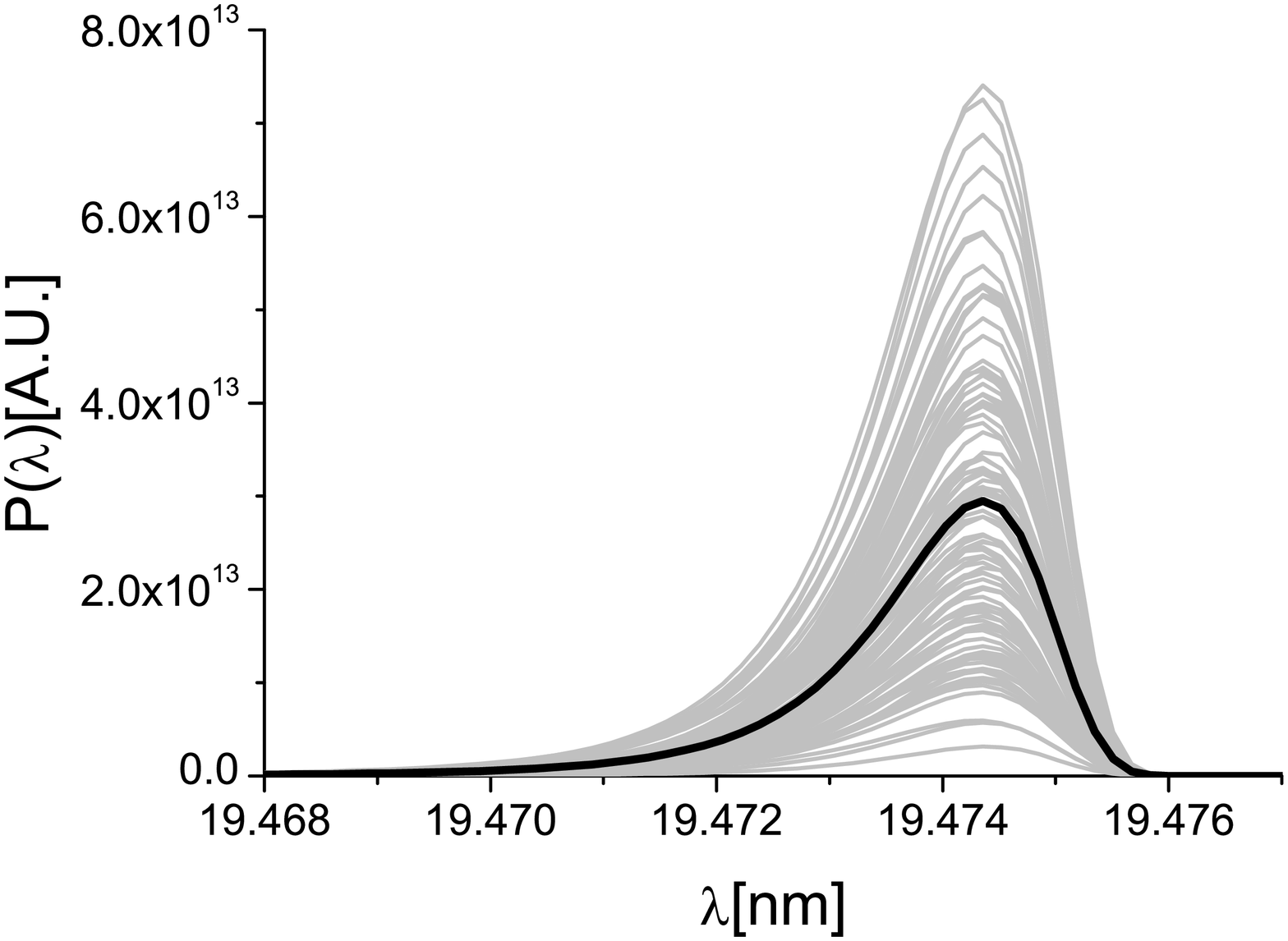}
\includegraphics[width=0.5\textwidth]{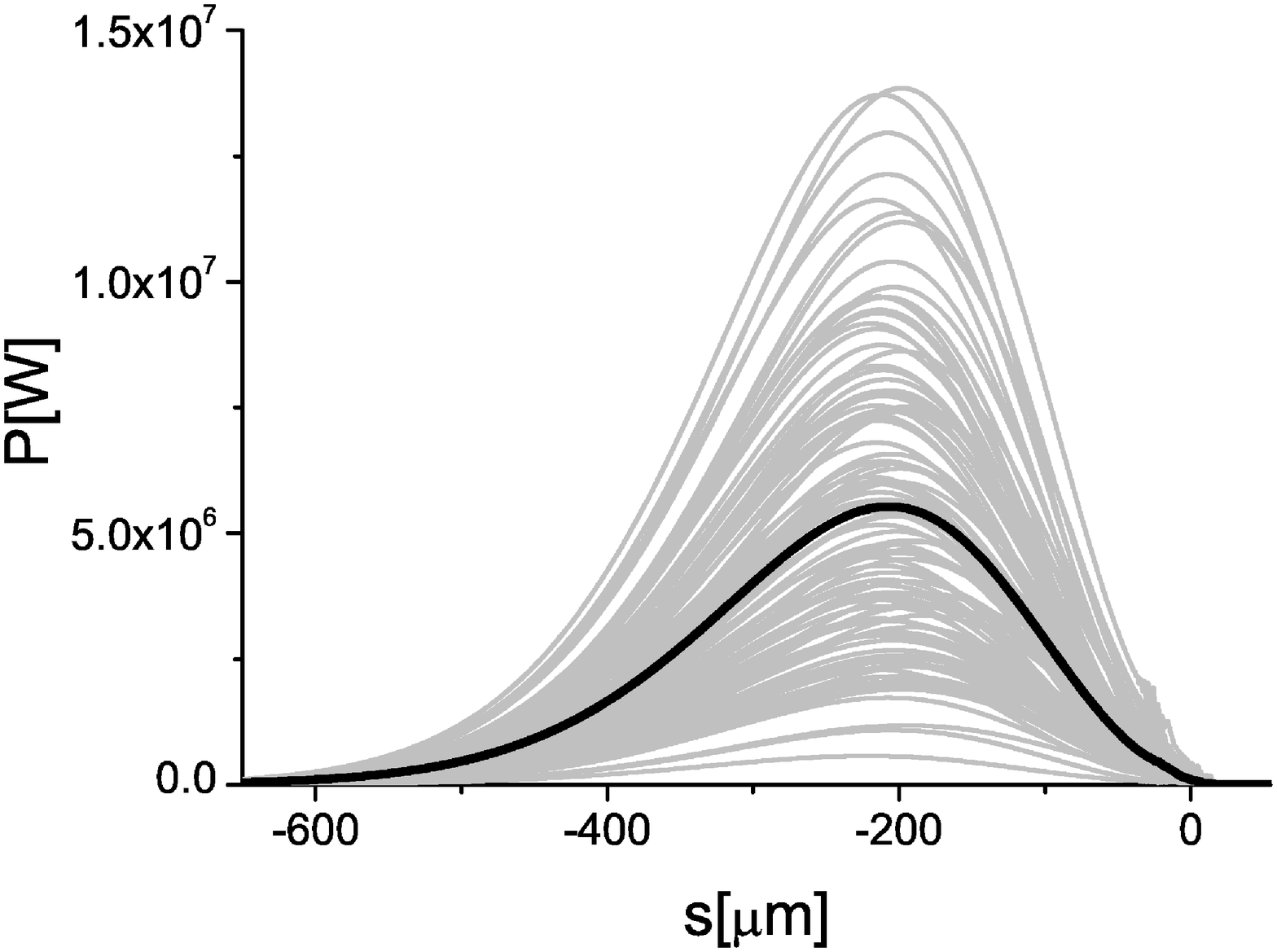}
\caption{Second resonance mode of operation. Pulse spectrum (left plot) and power (right plot) after the gas cell. Grey lines refer to single shot realizations, the black line refers to an average over one hundred realizations.} \label{OUT2}
\end{figure}
The effect of the filter is shown in Fig. \ref{OUT2}, where we plot the output spectra (left plot) and power (right plot). The shot-to-shot fluctuations in the profiles of both intensity and spectrum are almost absent (although, for a single mode, we have a power fluctuation near to $100 \%$). In fact, in this case the filter bandwidth is around $5.3$ meV FWHM, which is already much narrower than the bandwidth of a single SASE spike. As a result, we are approaching the single-mode case.

\subsection{Third resonance at $64.466$ eV}

We now turn to consider the third autoionization profile for $n=4$ for the $(sp,2n+)$ autoionizing series of Helium. FLASH is tuned at saturation, so that the SASE average bandwidth overlaps the third resonance. In this paragraph we focus on the transition between bandstop to passband filter, as the gas column density increases.

\begin{figure}[tb]
\includegraphics[width=0.5\textwidth]{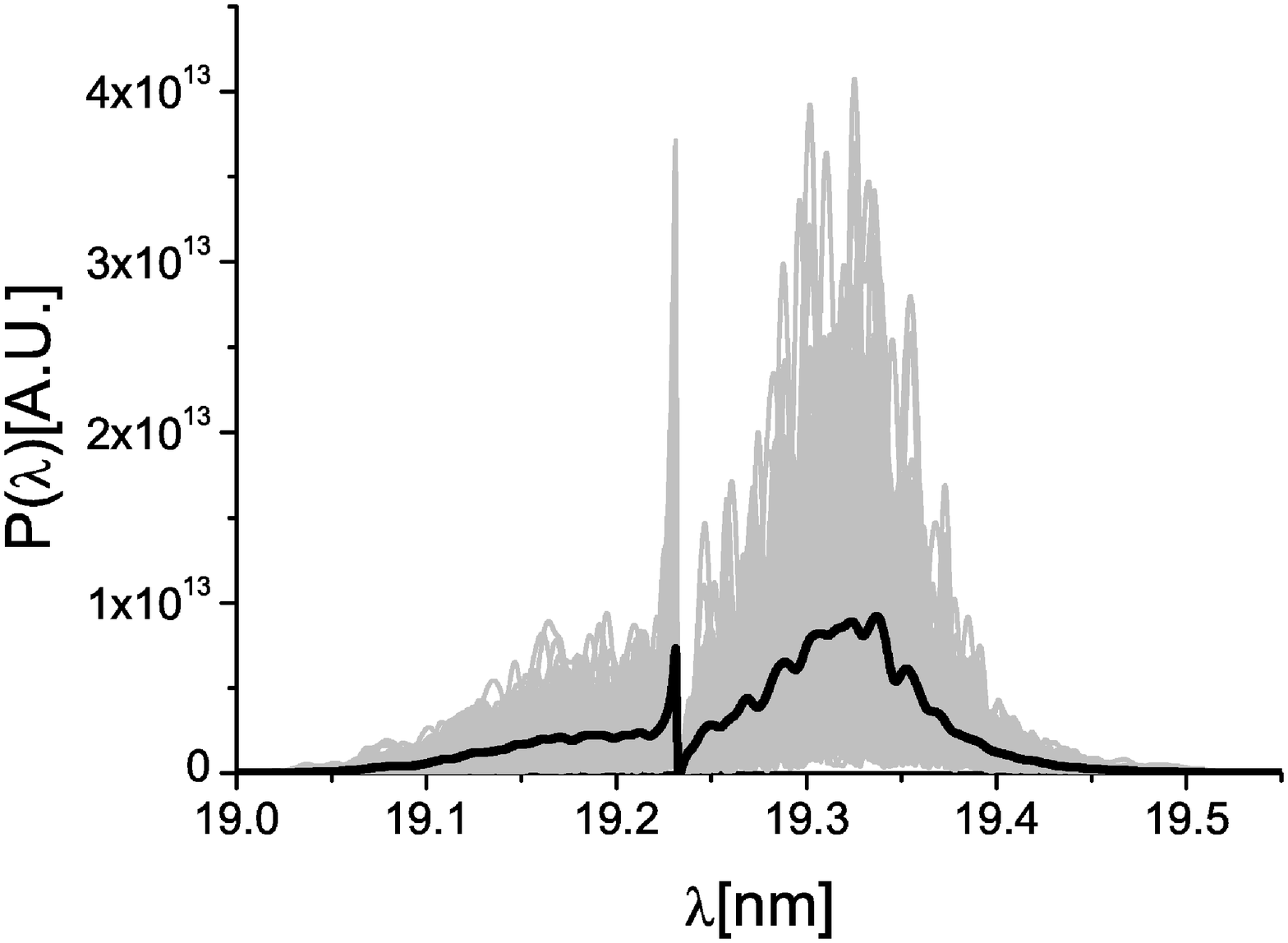}
\includegraphics[width=0.5\textwidth]{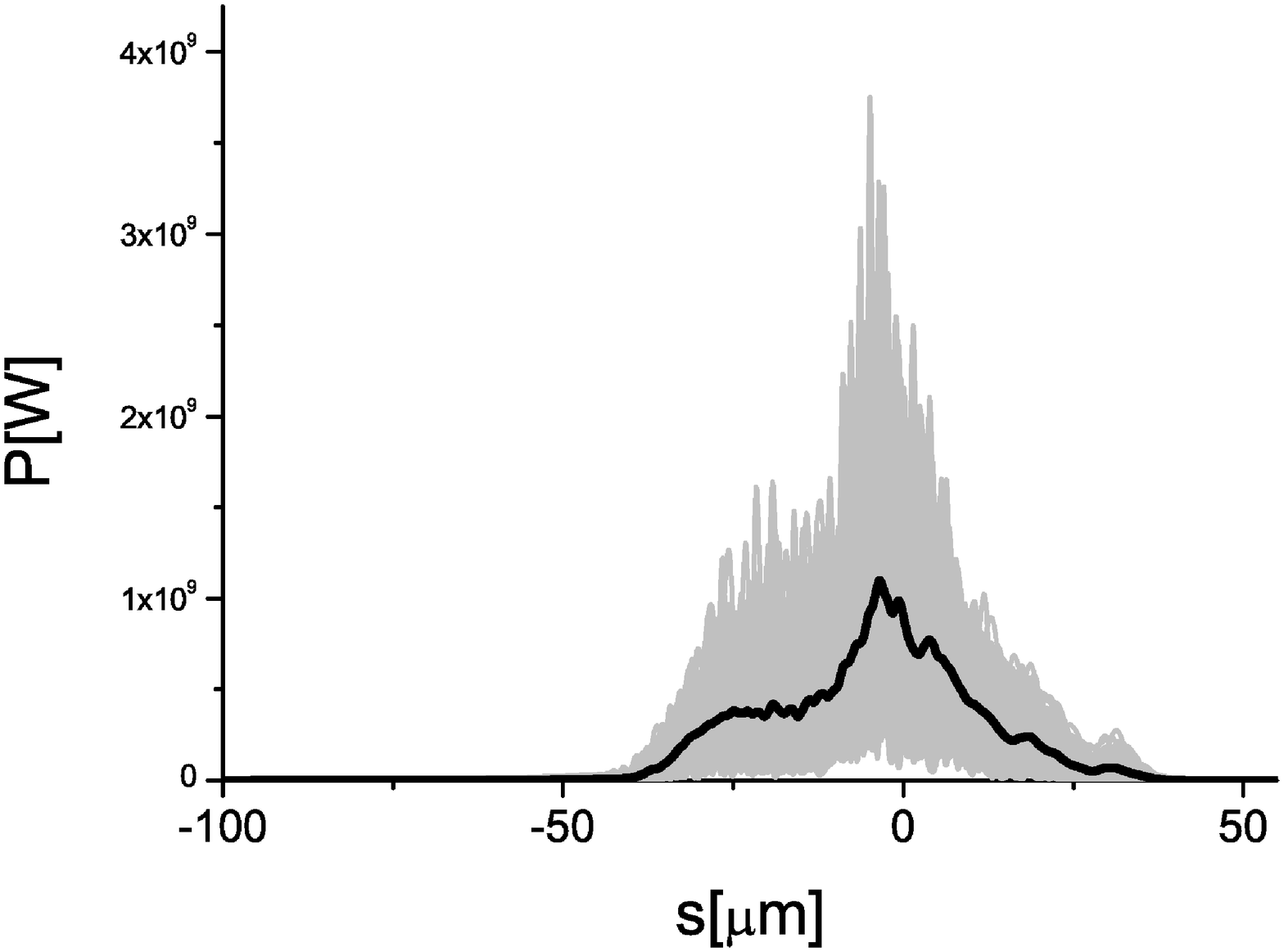}
\caption{Third resonance mode of operation. Pulse spectrum (left plot) and power (right plot) after the gas cell. Grey lines refer to single shot realizations, the black line refers to an average over one hundred realizations. Plots refer to a column density of $n_0 l = 10^{18} \mathrm{cm}^{-2}$.} \label{OUT3_1}
\end{figure}

\begin{figure}[tb]
\includegraphics[width=0.5\textwidth]{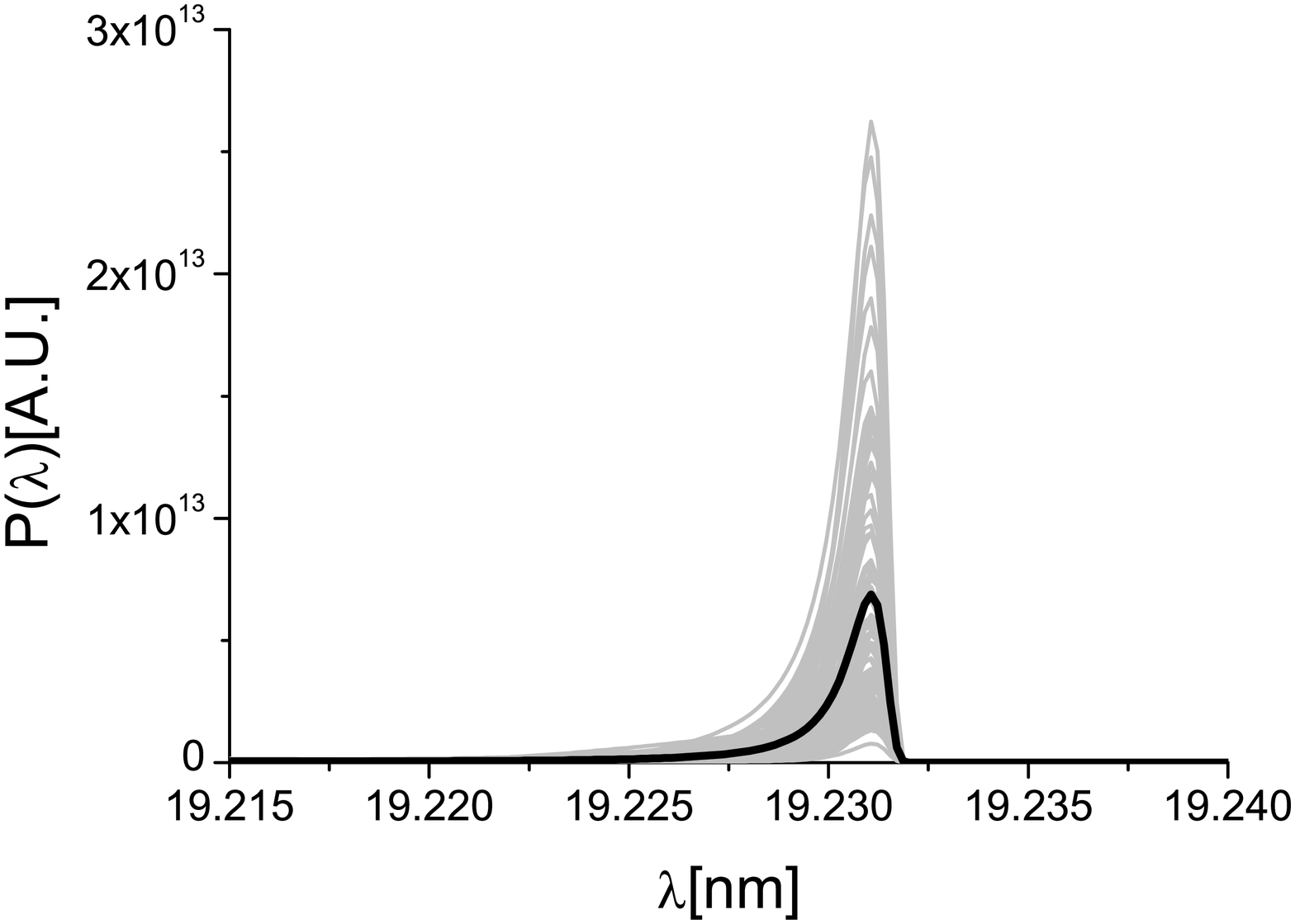}
\includegraphics[width=0.5\textwidth]{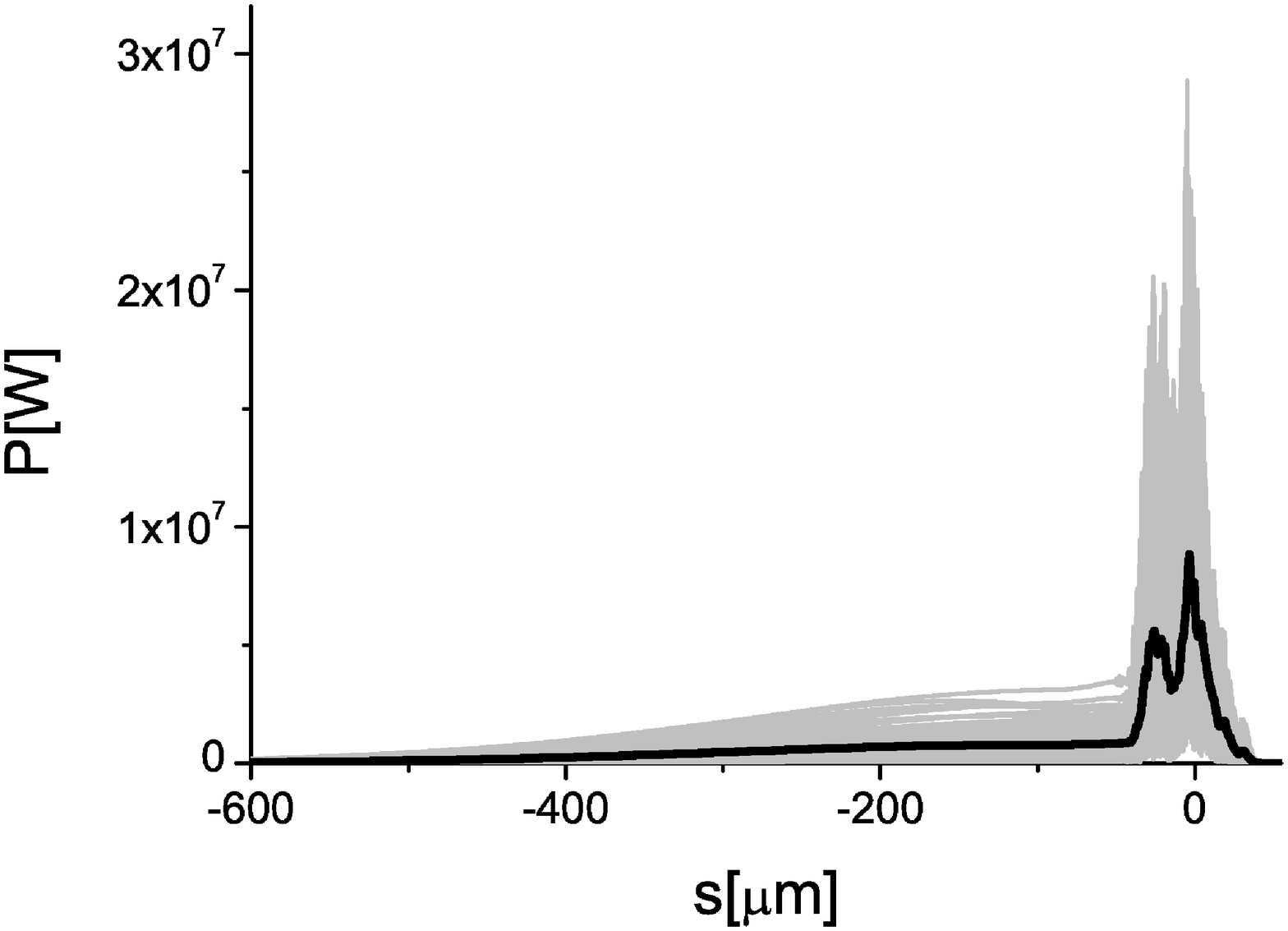}
\caption{Third resonance mode of operation. Pulse spectrum (left plot) and power (right plot) after the gas cell. Grey lines refer to single shot realizations, the black line refers to an average over one hundred realizations. Plots refer to a column density of $n_0 l = 5\cdot 10^{18} \mathrm{cm}^{-2}$.} \label{OUT3_5}
\end{figure}
\begin{figure}[tb]
\includegraphics[width=0.5\textwidth]{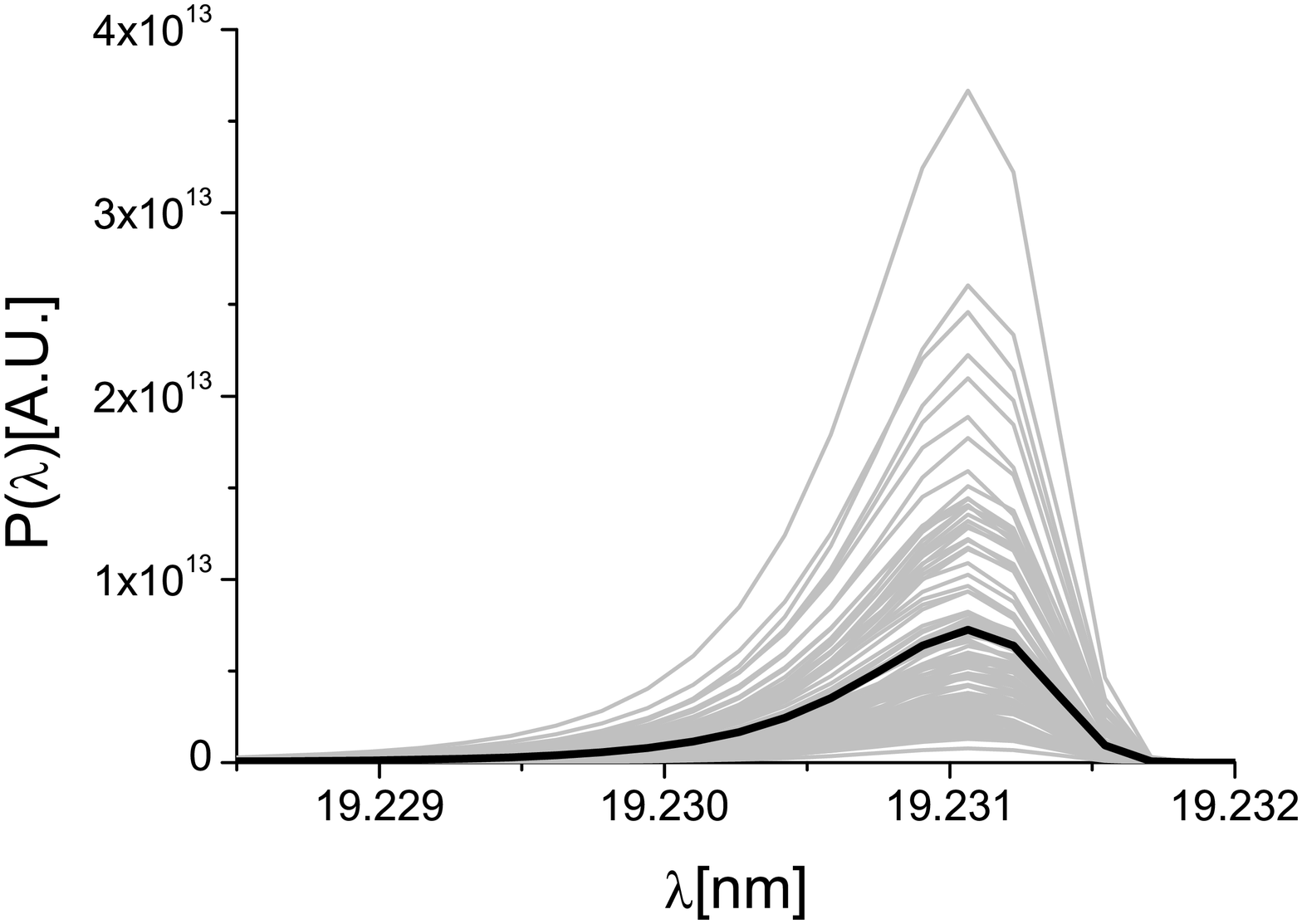}
\includegraphics[width=0.5\textwidth]{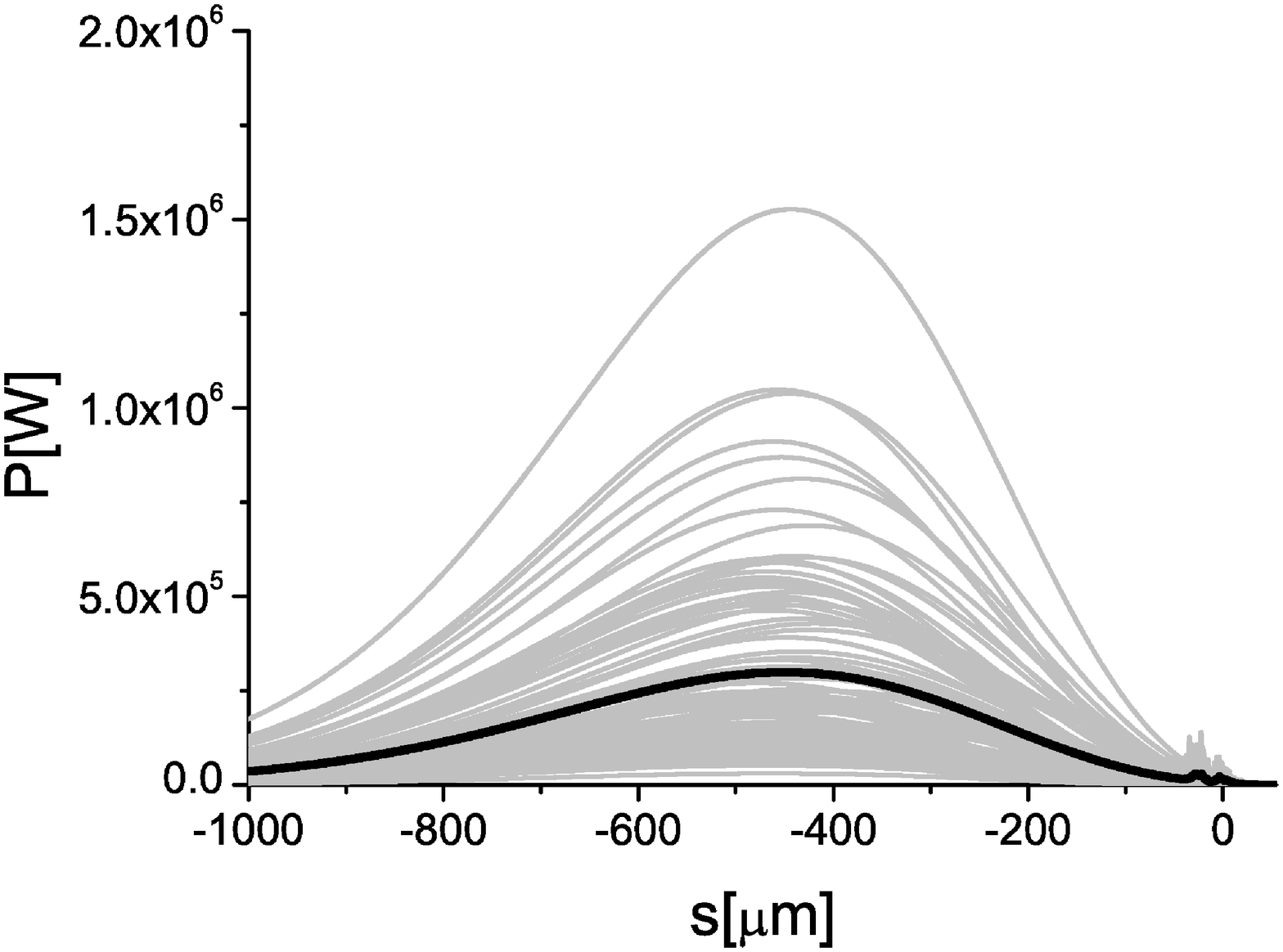}
\caption{Third resonance mode of operation. Pulse spectrum (left plot) and power (right plot) after the gas cell. Grey lines refer to single shot realizations, the black line refers to an average over one hundred realizations. Plots refer to a column density of $n_0 l = 10^{19} \mathrm{cm}^{-2}$.} \label{OUT3}
\end{figure}
\begin{figure}[tb]
\includegraphics[width=1.0\textwidth]{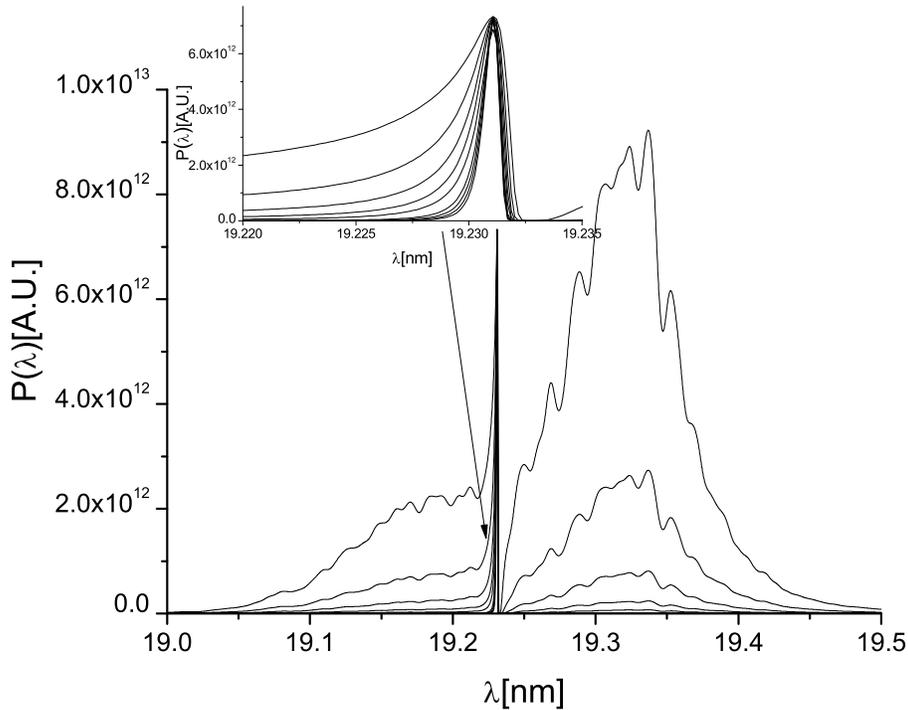}
\caption{Third resonance mode of operation. Pulse average spectra for a column density varying from $n_0 l = 10^{18} \mathrm{cm}^{-2}$ to $n_0 l = 10^{19} \mathrm{cm}^{-2}$ in steps of $n_0 l = 10^{18} \mathrm{cm}^{-2}$. The lines refers to  averages over one hundred realizations. The inset shows an enlargement of the bandpass range.} \label{specspec}
\end{figure}

As discussed before, when the gas pressure is decreased, the filter ceases to behave as a bandstop filter. This feature is shown by the comparison of Fig. \ref{OUT3_1}, Fig. \ref{OUT3_5} and Fig. \ref{OUT3}, which show, respectively, the output spectra and power for a column density of  $n_0 l =  10^{18} \mathrm{cm}^{-2}$,  $n_0 l = 5\cdot 10^{18} \mathrm{cm}^{-2}$ and $n_0 l =  10^{19} \mathrm{cm}^{-2}$. In Fig. \ref{specspec} we show the average power spectra for a column density varying from $n_0 l = 10^{18} \mathrm{cm}^{-2}$ to $n_0 l = 10^{19} \mathrm{cm}^{-2}$ in steps of $n_0 l = 10^{18} \mathrm{cm}^{-2}$.

In the case under study, the resonance width is only $2.1$ meV FWHM. For small gas column densities, the filter behaves as a bandstop filter as can be seen from Fig. \ref{specspec}. The inset of Fig. \ref{specspec}, or the left plot in Fig. \ref{OUT3} shows the output narrow bandwidth. As the gas pressure increases, we have an attenuation of the spectrum of many orders of magnitude, but the peak in the spectrum remains the same for all pressures. In other words, the peak efficiency of our monochromator is close to $100 \%$, the difference being only due to Doppler broadening, which, as we demonstrated before, constitutes a very small effect. For comparison, a conventional monochromator exhibits a peak efficiency of about $5\%-10\%$. As for the case of $n=3$, the shot-to-shot fluctuations in the profiles of both intensity and spectrum are almost absent (although, for a single mode, we have a power fluctuation near to $100 \%$), and we basically deal with a single-mode laser pulse.

\begin{figure}[tb]
\includegraphics[width=0.5\textwidth]{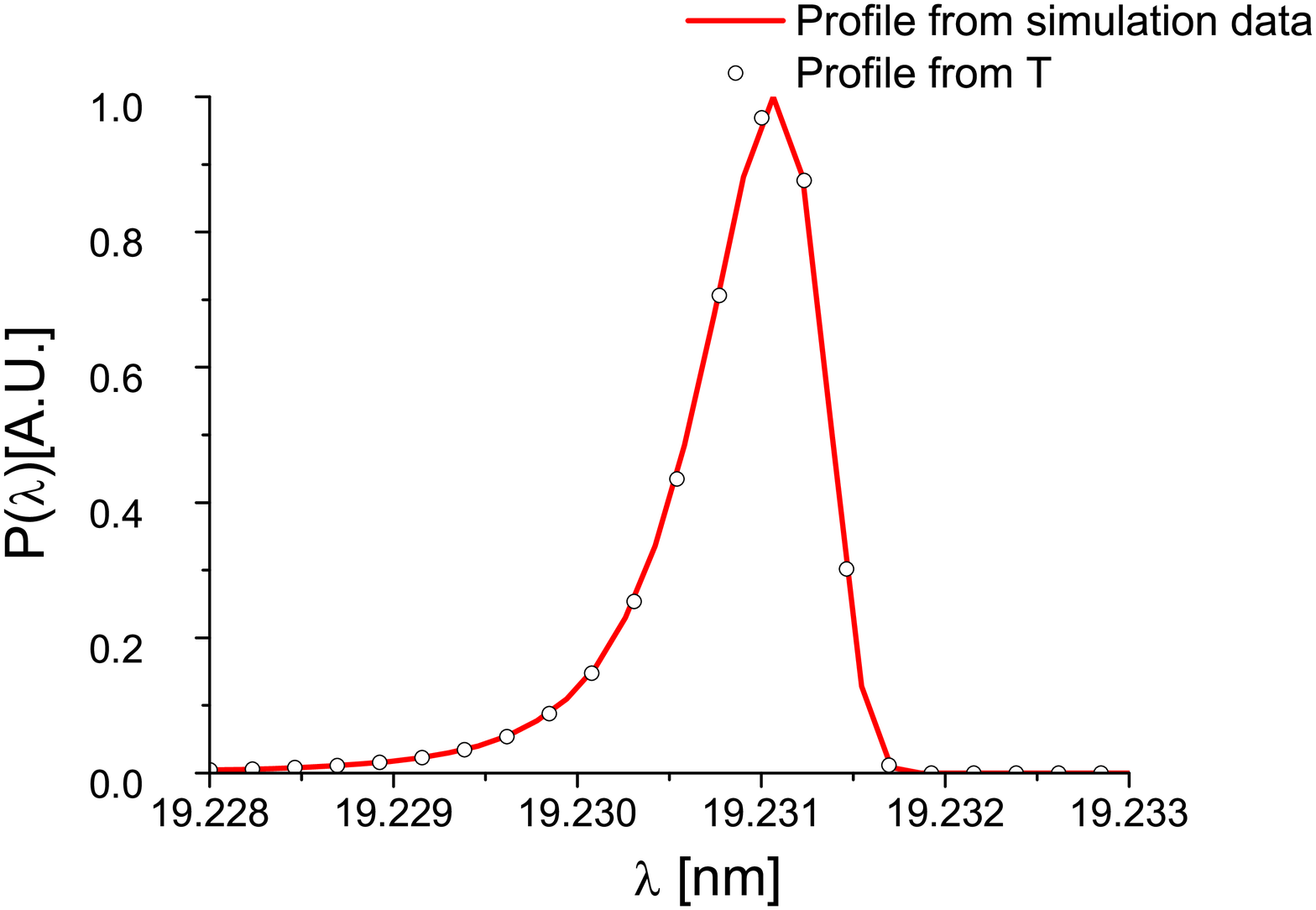}
\includegraphics[width=0.5\textwidth]{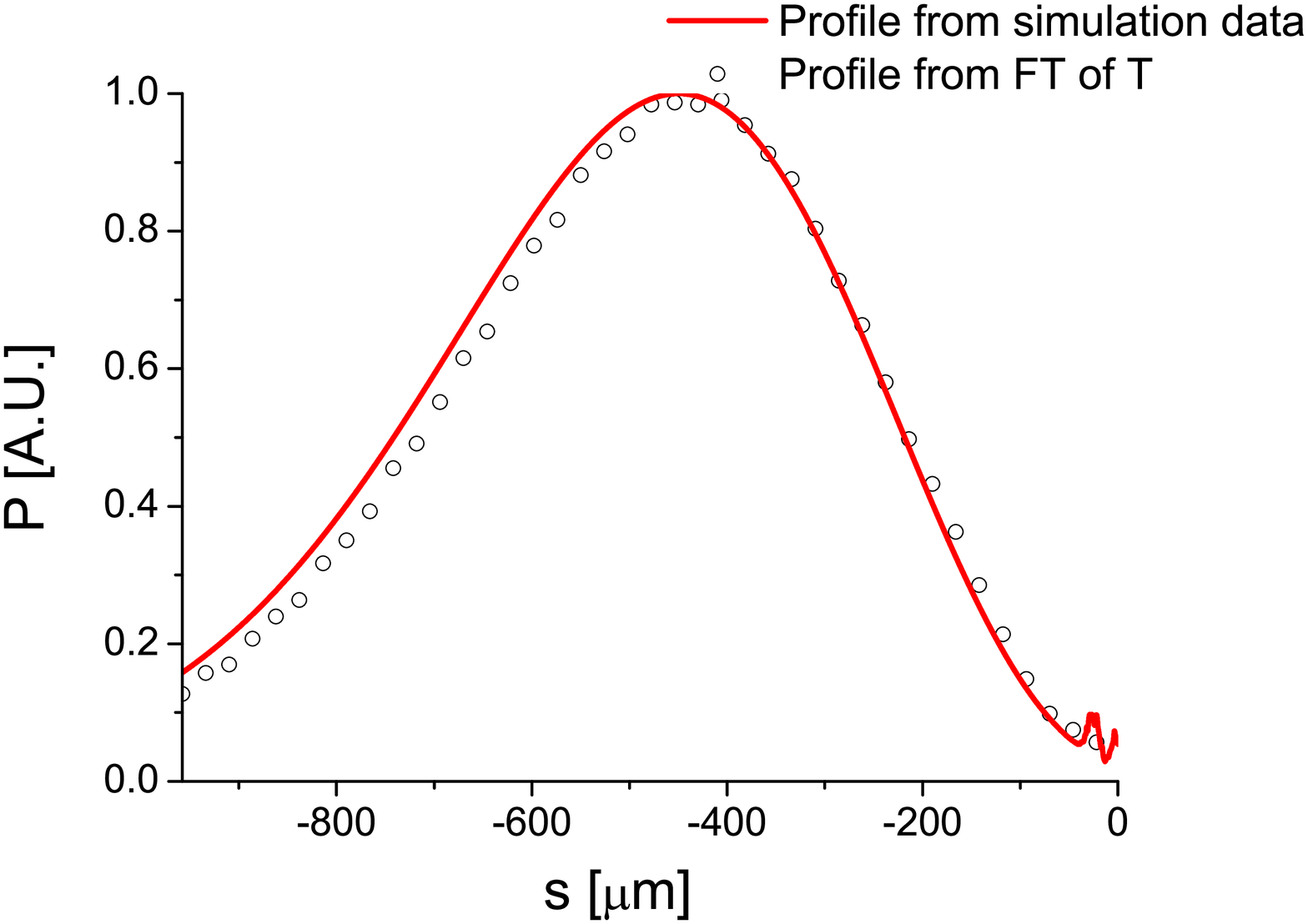}
\caption{Third resonance mode of operation. Left plot: comparison between the average spectrum in Fig. \ref{OUT3}, left, with the square modulus of the transmittance calculated with Eq. (\ref{ModT}) and Eq. (\ref{KKrel2}). Right plot: comparison between the average power profile in Fig. \ref{OUT3}, right, with the square modulus of the Fourier transform of the transmittance calculated with Eq. (\ref{ModT}) and Eq. (\ref{KKrel2}). All plots are normalized to unity. The gas column density is $n_0 l = 10^{19} \mathrm{cm}^{-2}$. } \label{specpow}
\end{figure}
Except for the amplitude, the characteristics of the single-mode pulse are fixed by the filter transmittance and not by the SASE spectra, which changes from shot to shot. This can be shown in the frequency domain by comparing, for example, the spectra in Fig. \ref{OUT3} with the square modulus of the right plot in Fig. \ref{transm}. In particular, we compared the average spectrum in Fig. \ref{OUT3}, left, with the square modulus of the transmittance calculated with Eq. (\ref{ModT}) and Eq. (\ref{KKrel2}) (and shown in Fig. \ref{transm}). The result is plotted in Fig. \ref{specpow}, left, where all plots are normalized to unity. It is also interesting to make such comparison in the time domain. This can be accomplished by comparing the right plot in Fig. \ref{OUT3} with the Fourier transform of the transmittance. In particular we compared the average power profile in Fig. \ref{OUT3}, right, with the square modulus of the Fourier transform of the transmittance calculated with Eq. (\ref{ModT}) and Eq. (\ref{KKrel2}).  The result is plotted in Fig. \ref{specpow}, right, where all plots are normalized to unity.

\subsection{Production of doublet spectral lines corresponding to the second and to the third resonance}

\begin{figure}[tb]
\includegraphics[width=1.0\textwidth]{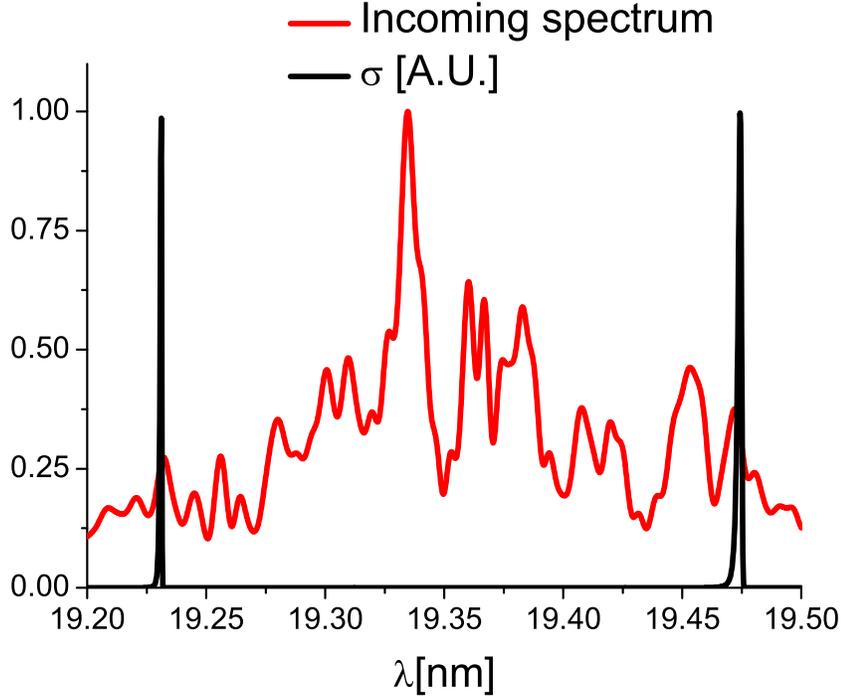}
\caption{The SASE spectrum after saturation overlaps the second and the third resonance. Here a single shot spectrum after six undulator modules is shown together with the Fano profiles relevant to the two resonances involved.} \label{Tsig4}
\end{figure}

Up to now, we considered cases where FLASH was tuned at saturation, so that the SASE average bandwidth overlapped with one resonance at a time, $n=2,3$ or $4$. However beyond saturation, the bandwidth of the SASE spectrum continues to increase, due to a deterioration of the coherence time. It is therefore possible to set the operating point of FLASH well after saturation, that is after six active undulator modules, corresponding to a magnetic length of $30$ m. In this way the SASE spectrum can be made overlap two (or, in principle, more) resonances, yielding a doublet (or, in principle, a multiplet) structure. The overlap between the spectrum of an incoming radiation pulse (single shot) and the relevant resonances for $n=3$ (second) and $n=4$ (third) are shown of Fig. \ref{Tsig4}.

\begin{figure}[tb]
\includegraphics[width=0.5\textwidth]{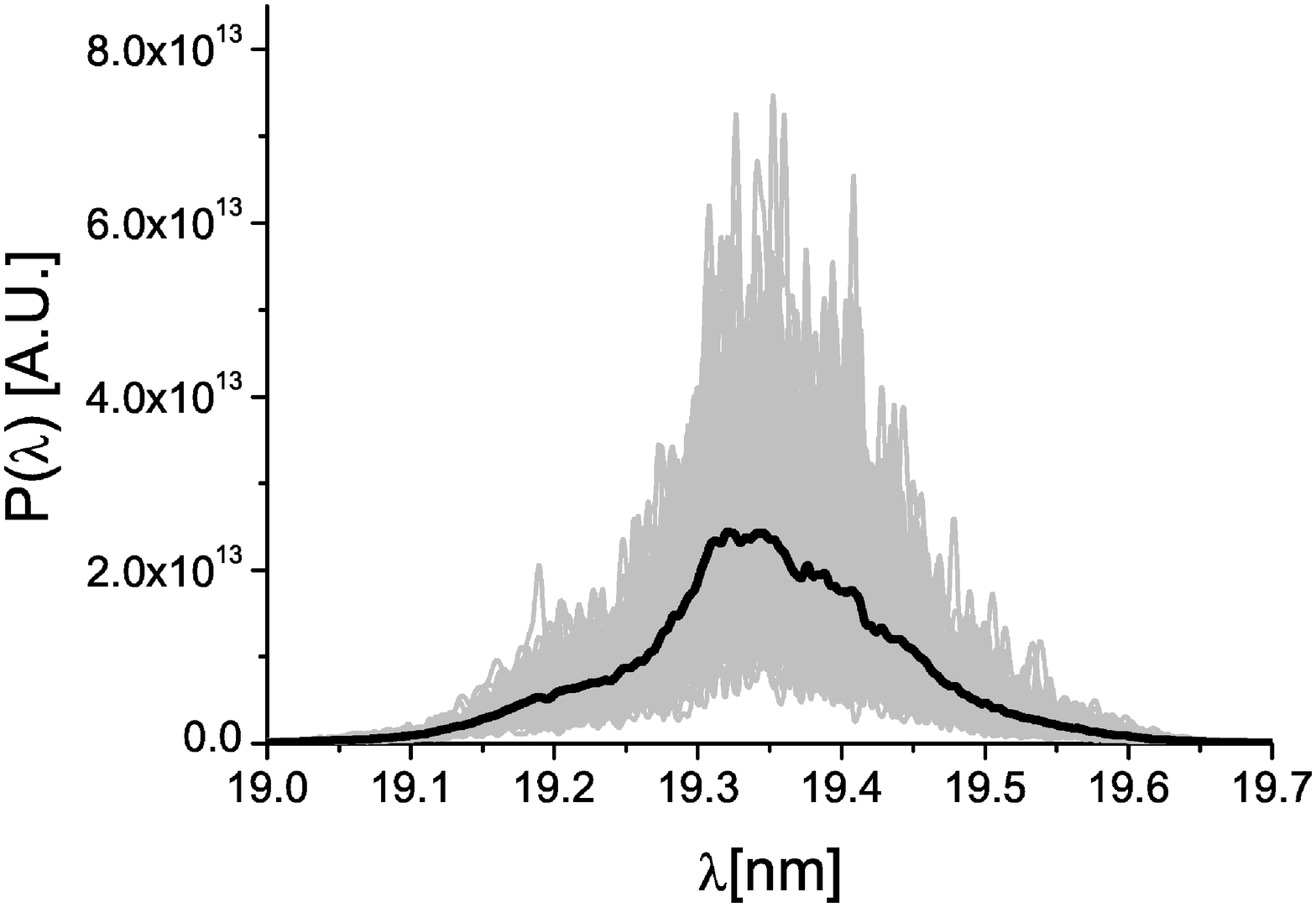}
\includegraphics[width=0.5\textwidth]{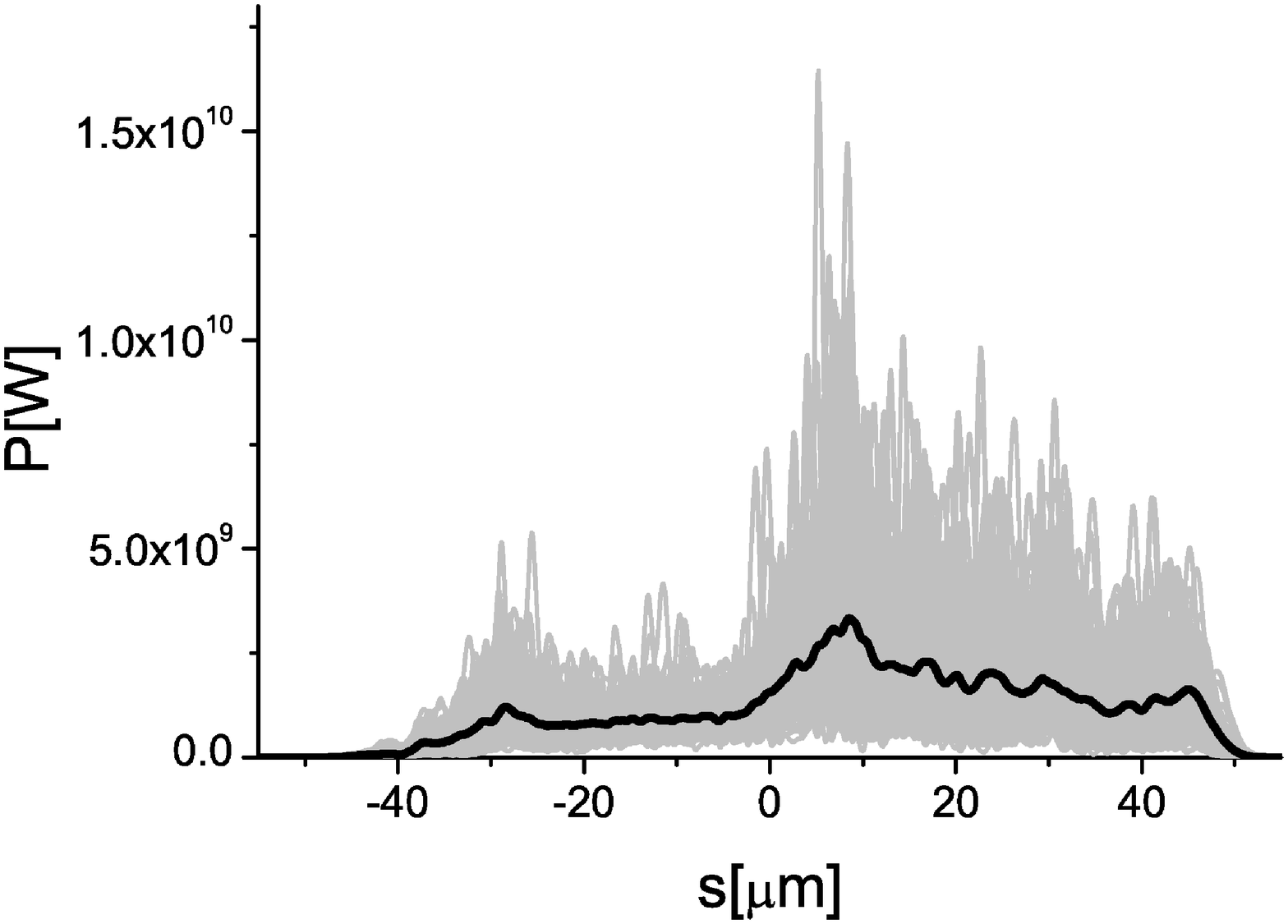}
\caption{Pulse spectrum (left plot) and power (right plot) before the gas cell. Grey lines refer to single shot realizations, the black line refers to an average over one hundred realizations.} \label{IN4}
\end{figure}
The input spectra and power are shown respectively in the left and right plots of Fig. \ref{IN4}.

\begin{figure}[tb]
\includegraphics[width=0.5\textwidth]{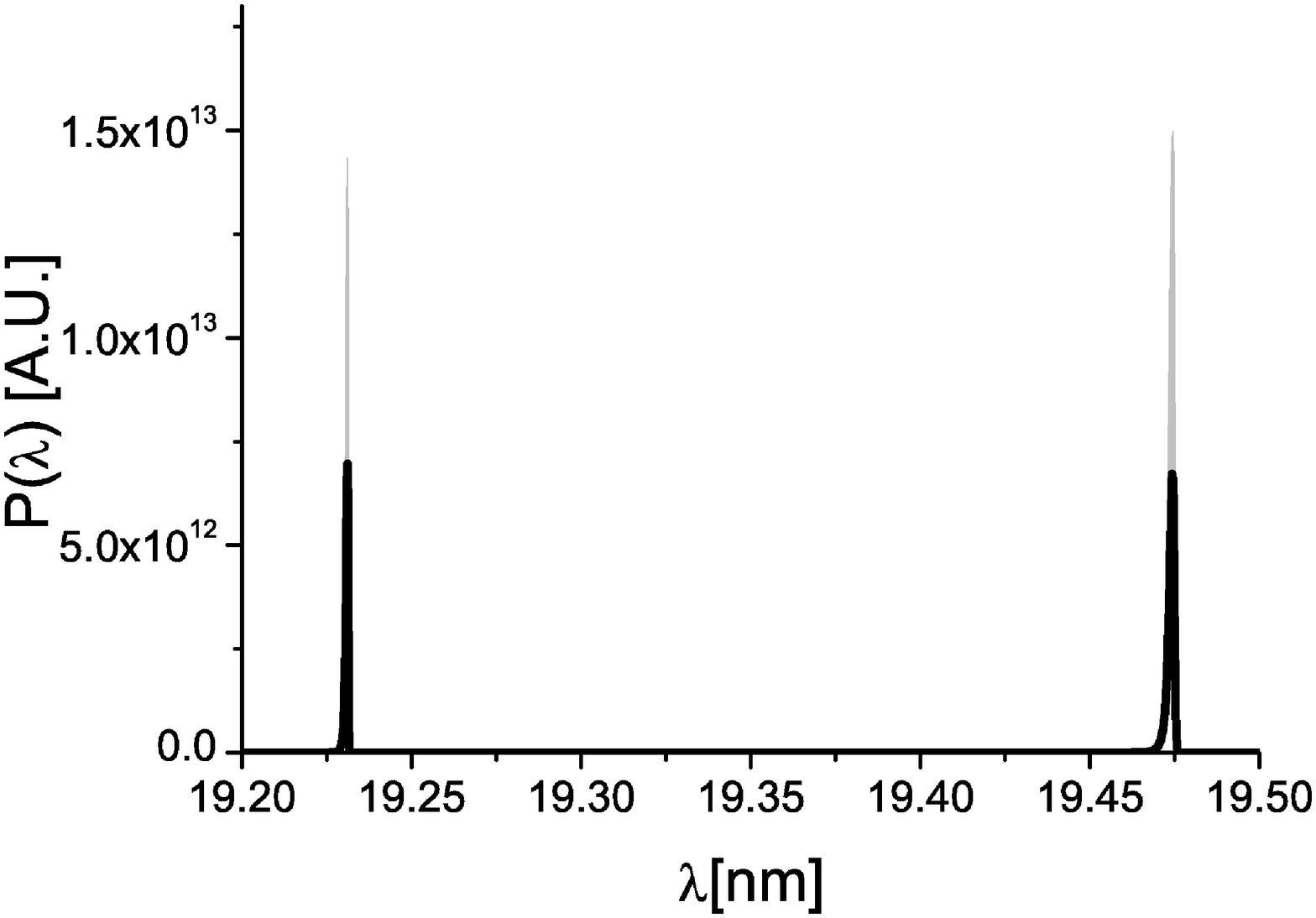}
\includegraphics[width=0.5\textwidth]{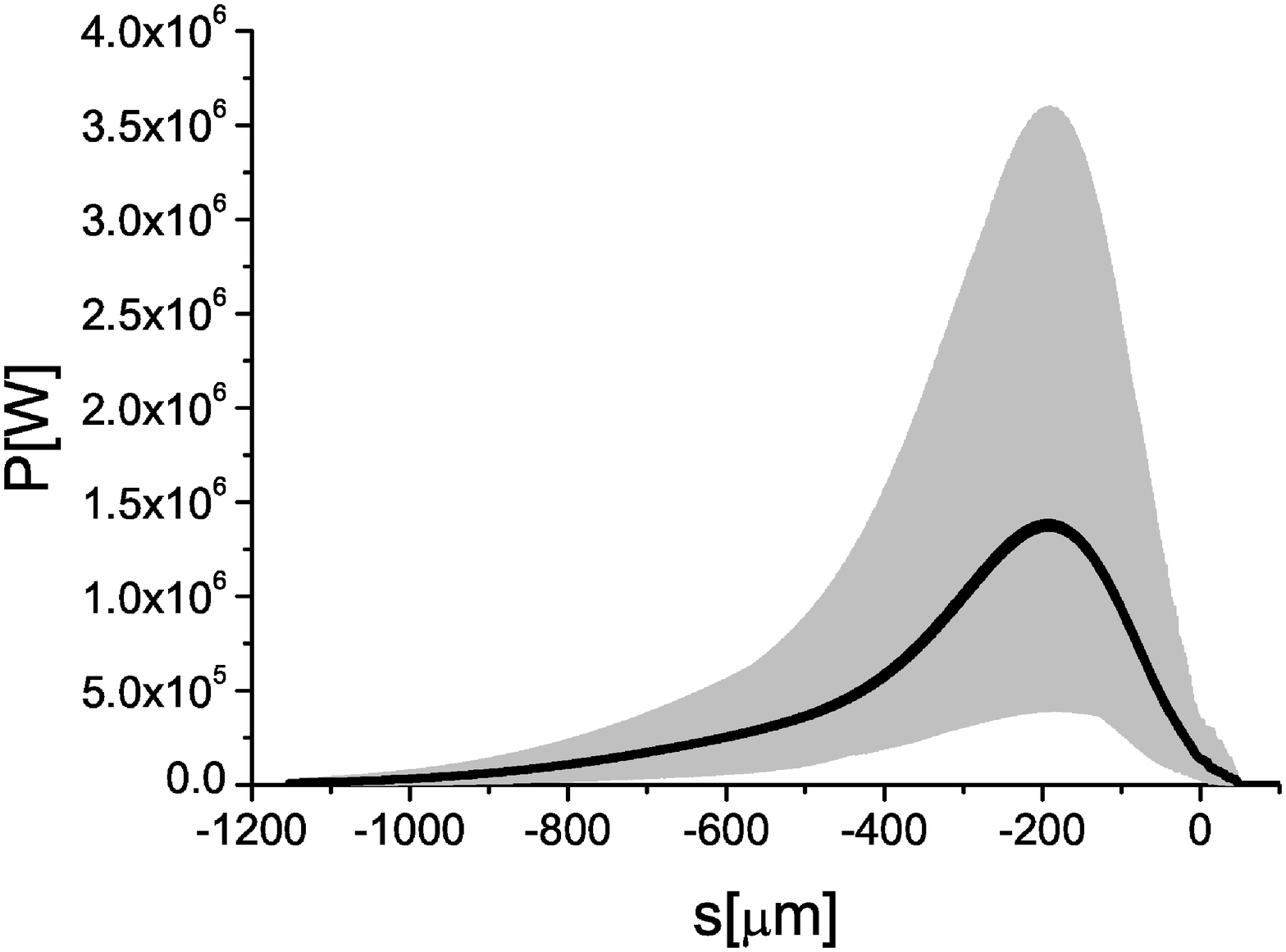}
\caption{Pulse spectrum (left plot) and power (right plot) after the gas cell. Grey lines refer to single shot realizations, the black line refers to an average over one hundred realizations.} \label{OUT4}
\end{figure}
\begin{figure}[tb]
\includegraphics[width=1.0\textwidth]{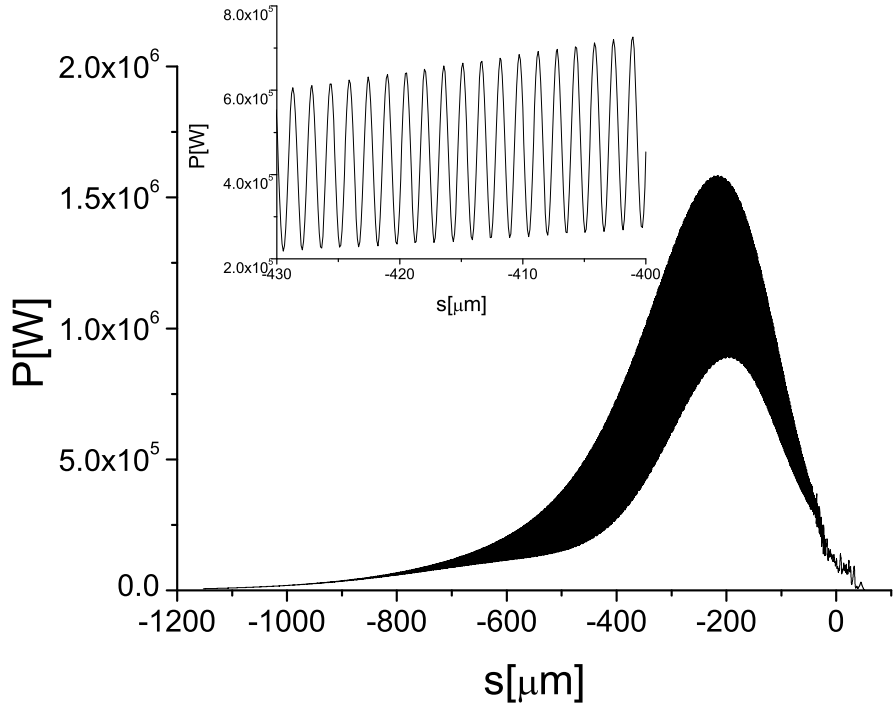}
\caption{Single-shot power distribution. The two frequency filters centered at different central frequencies are responsible for a beating between the two waves at slightly different frequencies, yielding (see inset) a power distribution modulated in the infrared range at a few microns wavelength. } \label{OUT4_S}
\end{figure}

The output spectra and power are shown respectively in the left and right plots in Fig. \ref{OUT4}. The effect of the superposition with the double resonance is clear from the analysis of the left plot. In is interesting to note that since there are now two frequency filters centered at different central frequencies, a beating between the two waves at slightly different frequencies takes place, yielding a power distribution modulated in the infrared range at a few microns wavelength. This effect is evident from the analysis of the inset of Fig. \ref{OUT4_S}, where we show the output power form a single shot pulse. This effect is similar to that considered in \cite{OURY6}. In that paper we were considering the possibility of self-seeding at two or more frequencies in the hard X-ray region, and the beating between the two waves was also present in that case.

\section{Conclusions}

We propose a new method to obtain a spectral filter in the VUV wavelength range using the Fano interference phenomenon for the VUV  photo absorption spectra of Helium. Our gas monochromator combines a very narrow bandwidth, down to a few meV and a very high peak efficiency of $99 \%$ with a
much-needed experimental simplicity. No optical elements nor alignment are required. The applicability of our scheme is limited to the wavelength range around $20$ nm. In this range, however it is possible to completely define the output VUV pulse characteristics, both in modulus and phase. This opens up the possibility of full characterization of the VUV radiation pulse in the time domain, which constitutes a powerful advantage for materials scientists wishing to model radiation-matter interactions.

\section*{Acknowledgements}

We are grateful to Massimo Altarelli, Reinhard Brinkmann, Serguei
Molodtsov and Edgar Weckert for their support and their interest
during the compilation of this work.

\end{document}